\documentclass[a4paper,10pt]{article}
\pdfoutput=1
\usepackage{jheppub}
\usepackage{mathtools}
\usepackage{physics}
\usepackage{multirow}
\usepackage{slashed}
\usepackage{graphicx}
\usepackage{hyperref}
\usepackage{tikz}

\allowdisplaybreaks  

\newcommand{\orcid}[1]{%
	\href{https://orcid.org/#1}{%
		\raisebox{-0.5ex}{%
			\includegraphics[height=1.8ex]{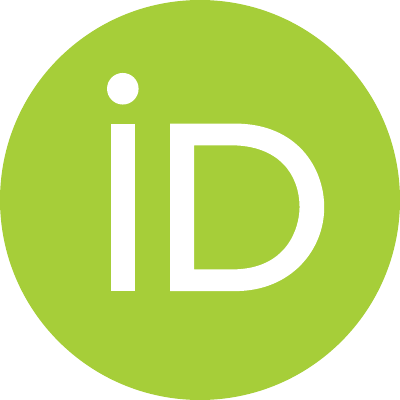}%
		}%
	}%
}

\title{\boldmath Probing the Electromagnetic Structure of Heavy-Light Hybrid Mesons with Vector and Axial-Vector Quantum Numbers}
\author[a,b]{A. Amiri\,\orcid{0000-0003-2479-4207},}
\emailAdd{amir.amiri1308@gmail.com}
\author[b,c]{K. Azizi\,\orcid{0000-0003-3741-2167},}
\emailAdd{kazem.azizi@ut.ac.ir}
\author[a]{P. Eslami\,\orcid{0000-0002-9308-2900}}
\emailAdd{eslami@um.ac.ir}

\affiliation[a]{Department of Physics, Faculty of Science, Ferdowsi University of Mashhad, P.O.Box 1436,\\ Mashhad, Iran}
\affiliation[b]{Department of Physics, University of Tehran, North Karegar Avenue, Tehran 14395-547,\\ Iran}
\affiliation[c]{Department of Physics, Dogus University, Dudullu-$\ddot{U}$mraniye, 34775 Istanbul, T$\ddot{u}$rkiye}

\abstract{We investigate the electromagnetic (EM) properties of vector and axial-vector heavy-light hybrid mesons within the light cone QCD sum rules framework. These states are characterized by an explicit gluonic degree of freedom and are studied through interpolating currents corresponding to the $J^{P(C)}=1^{-(-)}$, $1^{+(-)}$, $1^{+(+)}$, and $1^{-(+)}$ hybrid configurations. By analyzing the correlation function in the presence of an external background photon and matching its hadronic and QCD representations, we derive the light cone sum rules for the magnetic dipole and electric quadrupole form factors and extract the corresponding electromagnetic moments at $Q^2=0$. The QCD side of the sum rules incorporates the perturbative photon emission contributions as well as the relevant non-perturbative effects described by photon distribution amplitudes (DAs). Numerical predictions are obtained for the charged $\bar b g u$, $b g\bar u$, $\bar c g d$, $c g\bar d$, $\bar c g s$, and $c g\bar s$ hybrid configurations. The calculated electromagnetic moments show variations among the different hybrid configurations. A more systematic pattern is observed for the electric quadrupole moments (EQMs): the results obtained from the $\mathcal J_\mu^1$ and $\mathcal J_\mu^3$ currents are generally close to each other, as are those obtained from $\mathcal J_\mu^2$ and $\mathcal J_\mu^4$, with the former pair generally yielding somewhat larger magnitudes. The magnetic dipole moments (MDMs), in contrast, do not necessarily exhibit the same systematic grouping, although they also show variations among the different currents and configurations. For the bottom-containing states, particularly the $\bar b g u$ and $b g\bar u$ configurations, the quadrupole moments show a relatively weaker variation among the interpolating currents than those of the corresponding charm-containing states, while a similar tendency is less distinct for the magnetic dipole moments. The magnitudes of both electromagnetic moments are generally smaller for the bottom-containing states than for their corresponding charm-containing states. These results provide complementary information on the electromagnetic properties of heavy-light hybrid mesons and may be useful for comparisons among different hybrid configurations in future theoretical and experimental studies.
}

\begin{document} 
\begin{flushright}
\end{flushright}
\maketitle

\section{Introduction}
Quantum chromodynamics (QCD) provides the fundamental description of the strong interaction; yet the spectrum and internal structure of hadrons in the non-perturbative regime remain among the most challenging aspects of the theory. In the conventional quark model, mesons are predominantly described as color singlet quark-antiquark configurations, while gluons are responsible for mediating the interaction between the constituents. QCD, however, allows for a richer realization of hadronic matter in which gluonic degrees of freedom can participate explicitly in the hadron structure. Mesonic states containing a quark, an antiquark, and an excited gluonic degree of freedom are commonly referred to as hybrid mesons. Such states provide a particularly valuable laboratory for investigating the role of gluonic excitations in confinement and for testing the dynamics of QCD beyond the simplest quark model description~\cite{Jaffe:1975fd,Horn:1977rq,Dudek:2010wm,Dudek:2013yja,Meyer:2015eta}.

An important feature of hybrid mesons is that the additional gluonic degree of freedom enlarges the set of quantum numbers available to the meson spectrum. In particular, some $J^{PC}$ combinations, such as $1^{-+}$, cannot be constructed from a conventional quark-antiquark pair and are therefore referred to as spin exotic quantum numbers. Lattice QCD calculations have provided strong evidence for multiplets of hybrid states containing both exotic and non-exotic quantum numbers, with the gluonic excitation coupled to a color octet quark-antiquark pair~\cite{Dudek:2010wm,Dudek:2013yja}. Complementary descriptions have also been developed within flux tube, constituent gluon, bag model, and other approaches, emphasizing different realizations of the gluonic excitation~\cite{Meyer:2015eta}. The coexistence of exotic and non-exotic hybrid configurations is particularly important phenomenologically: while exotic quantum numbers provide a relatively clean signature beyond the conventional quark model, non-exotic hybrids can mix with ordinary $q\bar q$ states and are consequently much more difficult to identify experimentally.

The experimental situation has developed considerably in recent years. In the light sector, the long standing $\pi_1$ structures have provided important candidates for exotic hybrid mesons, with the $\pi_1(1600)$ receiving support from theoretical and experimental investigations~\cite{Meyer:2015eta,Chung:2002pu,Baker:2003jh,E852:2004rfa,COMPASS:2014vkj,JPAC:2018zyd,Kopf:2020yoa}. Recent experimental and theoretical studies, including analyses by GlueX, continue to investigate the $\pi_1(1600)$ and its possible decay and production channels~\cite{Woss:2020ayi,GlueX:2024erj}. The observation of the iso-scalar $\eta_1(1855)$ with exotic $J^{PC}=1^{-+}$ quantum numbers by the BESIII Collaboration has provided an additional important development in the search for hybrid mesons~\cite{BESIII:2022riz,BESIII:2022iwi}. A theoretical study identifies this state as a member of the $I = 0$ hybrid multiplets and shows that observing a single $\eta_1(1855)$ in the $\eta\eta'$ channel can provide a strong constraint on the hybrid scenario~\cite{Qiu:2022ktc}. Recently, the COMPASS Collaboration reported evidence for a supernumerary pseudo-scalar resonance, denoted as $K(1690)$, in the $K^-\pi^+\pi^-$ invariant mass spectrum produced in the scattering reaction $K^-p\rightarrow K^-\pi^+\pi^-p$, with a mass of approximately $1.7~\mathrm{GeV}/c^2$~\cite{COMPASS:2025wkw}. The state was interpreted by the Collaboration as a crypto-exotic strange meson with quantum numbers $J^P=0^-$. More recently, a theoretical investigation of its mass spectrum and strong decay properties found that the observed characteristics are compatible with a hybrid meson interpretation of the $K(1690)$~\cite{Chen:2026ets}. These developments demonstrate both the progress in the experimental search for hybrid mesons and the continuing need for observables capable of distinguishing their internal structure from that of conventional or other exotic configurations.

The heavy quark sectors provide another important arena for the study of gluonic excitations. Hidden heavy hybrid mesons, particularly $c g \bar c$, $b g \bar b$, and $\bar b g c$ states, have been extensively investigated using lattice QCD, QCD sum rules, and other approaches~\cite{Chen:2013zia,Miyamoto:2019oin,Alaakol:2024zyh,Barsbay:2025vjq,Schlosser:2025tca,Shuryak:2026yoi}. On the experimental side, several states in the XYZ spectrum have been discussed as possible candidates for states with non-conventional structures. In particular, the state originally identified as $Y(4260)$ was proposed early on as a possible charmonium hybrid, motivated by its mass and decay properties~\cite{Miyamoto:2019oin,BaBar:2005hhc}. However, its interpretation remains unsettled, with molecular, tetraquark, conventional charmonium, and hybrid scenarios having all been investigated~\cite{Miyamoto:2019oin}. More recent theoretical analyses continue to investigate possible hybrid assignments of heavy quarkonium states and their radiative and hadronic transitions. Lattice calculations have also demonstrated that EM transition form factors provide valuable information about the internal structure of heavy quark systems and can, in principle, be used to investigate higher-lying and exotic states~\cite{Dudek:2006ej,Delaney:2023fsc}. It is important, however, to distinguish the hidden heavy sector from heavy-light hybrids. The latter have received considerably less theoretical attention, emphasizing the importance of reliable theoretical predictions for their masses and EM properties.

Heavy-light hybrids, consisting of a single heavy quark, a light quark, and an explicit gluonic excitation, constitute a complementary class of exotic hadrons. Lattice studies of highly excited charm-light and charm-strange mesons have identified candidate hybrid states in the $D$ and $D_s$ spectra through their strong overlap with operators proportional to the gluonic field strength tensor~\cite{Moir:2012lxe,Cheung:2016bym}, while similar hybrid supermultiplet patterns have also been explored in the bottom-light sector, with identification of such hybrid states with different quantum numbers~\cite{Gayer:2024akw}. The continuing experimental exploration of open-flavor hadrons, including the observed $X(5568)$ structure, has further stimulated interest in this sector~\cite{D0:2016mwd,LHCb:2016dxl}. Since heavy-light hybrid states are not eigenstates of charge conjugation, they are characterized by non-exotic $J^P$ quantum numbers. Complementary QCD sum-rule studies have investigated the masses and current couplings of heavy-light hybrids with different spin-parity assignments and quark contents using interpolating currents involving the gluon field strength and its dual~\cite{Ho:2016owu,Barsbay:2022gtu}. These studies indicate that a rich spectrum of heavy-light hybrid configurations is possible in both the charm and bottom sectors. Nevertheless, their EM structure remains considerably less explored than their masses.

EM observables offer a particularly useful complementary probe of the internal structure of hadrons. While masses and strong decay widths provide important information about the spectrum and possible decay mechanisms, EM form factors directly probe the response of a hadronic state to an external EM field. For a spin-one state, the EM vertex contains several independent form factors whose values at the real photon point determine, in particular, the electric charge, MDM, and EQM. The MDM is sensitive to the spin and magnetization distributions of the constituents, whereas the EQM probes the deviation from spherical symmetry in the charge distribution. Consequently, these observables can carry information about the internal organization of the quarks and gluonic degrees of freedom that is not necessarily apparent from the mass spectrum alone. Lattice studies of EM and radiative form factors have demonstrated the usefulness of such quantities in probing the structure of heavy hadrons, while QCD light cone sum rules have provided a systematic framework for studying EM moments of composite hadronic states~\cite{Delaney:2023fsc,Dudek:2006ej,Ball:2002ps,Ozdem:2022ydv,Azizi:2023gzv,Ozdem:2026wmf}. 

The light cone QCD sum rule method is particularly suitable for this purpose because it provides a framework in which the perturbative photon emission dynamics and the non-perturbative photon structure can be treated within a common QCD based approach. In the presence of an external real photon, the long distance emission of the photon is described in terms of photon distribution amplitudes of increasing twist. A systematic classification of these DAs up to twist four was developed by Ball and collaborators, providing the non-perturbative input required for a wide range of light cone QCD sum rule calculations involving real photons~\cite{Ball:2002ps}. The method has subsequently been applied extensively to EM form factors and magnetic and quadrupole moments of various conventional and exotic hadrons~\cite{Azizi:2009egn,Aliev:2009gj,Aliev:2009np,Aliev:2019lsd,Azizi:2015ksa,Ozdem:2024qaa,Ozdem:2017exj,Ozdem:2017jqh,Azizi:2018mte,Ozdem:2018qeh,Ozdem:2024lpk,Ozdem:2025ncd}. Such applications demonstrate that EM moments can be sensitive to the internal configuration assumed for an exotic state and can therefore complement spectroscopic analyses.

Despite this progress, the EM properties of heavy-light hybrid mesons have received considerably less attention, particularly for vector and axial-vector states. This gap is important for several reasons. First, vector and axial-vector hybrids provide a natural opportunity to investigate the EM response of spin one states in the presence of an explicit gluonic excitation. Second, the availability of multiple interpolating currents with different quark-gluon configurations for the same quantum numbers allows one to study how the EM response depends on the internal structure of the hybrid state, making these observables potentially useful for distinguishing different hybrid configurations. Third, comparing states with identical flavor content but different hybrid configurations allows one to examine directly how the gluonic and spin structures influence EM observables. Finally, comparing the charm and bottom sectors provides an opportunity to investigate how the EM response evolves with the heavy quark mass and to explore qualitative aspects of the heavy quark limit.

Motivated by these considerations, in the present work we investigate the MDMs and EQMs of charged vector and axial-vector heavy-light hybrid mesons within the framework of light cone QCD sum rules~\cite{Khodjamirian:1995uc,Colangelo:2000dp,Bijnens:2002mg}. We consider the configurations with quark contents $\bar b g u$, $b g\bar u$, $\bar c g d$, $c g\bar d$, $\bar c g s$, and $c g\bar s$, and employ the set of interpolating currents that describe the $J^{P(C)}=1^{-(-)}$, $1^{+(-)}$, $1^{+(+)}$, and $1^{-(+)}$ hybrid configurations. The correlation functions are calculated by consistently incorporating the perturbative and non-perturbative photon emission contributions, including the relevant photon DAs. After matching the hadronic and QCD representations, double Borel transformation and continuum subtraction are employed to suppress the contributions of higher resonances and the hadronic continuum and to isolate the ground state contributions. The resulting light cone sum rules are then used to determine the MDMs and EQMs of the considered hybrid states.

The main purpose of this study is not only to provide numerical predictions for these electromagnetic observables but also to investigate their sensitivity to the internal quark-gluon configuration and to the heavy quark flavor. By comparing different interpolating configurations with the same quark and gluon content, as well as corresponding charm and bottom containing states, we aim to determine whether EM moments can provide useful structural information about heavy-light hybrids. The results may therefore serve as complementary theoretical input for future spectroscopic and experimental investigations of heavy-light hybrid mesons and contribute to a more complete understanding of the role of gluonic degrees of freedom in the hadron spectrum.

The remainder of this paper is structured as follows. In Section~\ref{sec:model}, we develop the theoretical framework employed in this work. The correlation function is formulated within the light cone QCD sum rule and background field approaches, followed by the construction of its hadronic and QCD representations for the vector and axial-vector hybrid states. The corresponding light cone sum rules for the MDMs and EQMs are then derived. 
The numerical results are discussed in Section~\ref{sec:Num}, where we examine the pole contribution, OPE convergence, and stability of the physical observables, together with the predictions for the MDMs and EQMs of the considered hybrid mesons.
Section~\ref{sec:conclusions} provides a summary of the main findings of this work.
Finally, Appendix~\ref{AppenPDAs} provides the photon DAs along with their relevant parameters, whereas the QCD representations of the correlation functions are presented in Appendix~\ref{AppenCorQCD}.
\section{Theoretical framework}\label{sec:model}
The theoretical framework underlying the analysis is formulated in this section. The correlation function is first introduced within the light cone QCD sum rule and background field approaches. Its hadronic and QCD representations are subsequently constructed for the vector and axial-vector hybrid mesons, providing the basis for deriving the corresponding light cone sum rules for their MDMs and EQMs.
\subsection{Correlation function}
The MDM ($\mu$) and EQM ($\mathcal{D}$) of the vector and axial-vector hybrid mesons are determined from the following three point correlation function, constructed within the light cone QCD sum rule framework:
\begin{equation}
	\label{edmn01}
	\Pi _{\mu \nu \alpha }(p,q)=i^2\int d^{4}x\,\int d^{4}y\,e^{ip\cdot x+iq \cdot y}\,
	\langle 0|\mathcal{T}\{\mathcal{J}_{\mu}(x) \mathcal{J}_{\alpha}^{em}(y)
	\mathcal{J}_{\nu }^{\dagger }(0)\}|0\rangle\,,  
\end{equation}%
where $\mathcal{J}_{\alpha}^{em}$ denotes the EM current, whereas $\mathcal{J}_{\mu(\nu)}$ represents the interpolating current associated with a vector or axial-vector hybrid state. The interaction of the photon with the hybrid meson is encoded in the matrix element $\langle \rm HM_{V(AV)}(p)|\mathcal{J}_{\alpha}^{em}|\rm HM_{V(AV)}(p')\rangle$, which enters the correlation function in Eq.~(\ref{edmn01}) for the initial and final states of the same hybrid meson. Here, $p'=p+q$ and $p$ denoting the momenta of the initial and final hybrid mesons, respectively. The EM current takes the form,
\begin{align}
	\label{edmn02}
	\mathcal{J}_{\alpha}^{em}(x) =\sum_{q= u,d,s,c,b} e_q \bar q(x) \gamma_\alpha q(x)\,.
\end{align}
where, $e_q$ represents the electric charge associated with the quark $q$. The interpolating currents for the hybrid meson states are chosen as,
\begin{eqnarray}
	\label{edmn03}
	&&\mathcal{J}^{1}_{\mu}(x)={g_s} \overline{Q}^{a}(x)\gamma^{\theta}\gamma_{5} \frac{\lambda _{ab}^{n}}{2} \tilde{G}_{\mu\theta}^{n}(x)q^{b} (x)\,,  \label{eq:Curr1}
\end{eqnarray}%
which couples to hybrid states with $0^{+(-)}$ and $1^{-(-)}$ quantum numbers,
\begin{eqnarray}
	\label{edmn04}
	&&\mathcal{J}_{\mu }^{2}(x)=g_s \overline{Q}^{a}(x)\gamma^{\theta}\gamma _{5} \frac{\lambda _{ab}^{n}}{2} G_{\mu\theta}^{n}(x)q^{b} (x)\,,  \label{eq:Curr2}
\end{eqnarray}%
which couples to hybrid states with $0^{-(-)}$ and $1^{+(-)}$ quantum numbers,
\begin{eqnarray}
	\label{edmn05}
	&&\mathcal{J}_{\mu }^{3}(x)={g_s} \overline{Q}^{a}(x)\gamma^{\theta} \frac{\lambda _{ab}^{n}}{2} \tilde{G}_{\mu\theta}^{n}(x)q^{b} (x)\,,  \label{eq:Curr3}
\end{eqnarray}%
which couples to hybrid states with $0^{-(+)}$ and $1^{+(+)}$ quantum numbers, and
\begin{eqnarray}
	\label{edmn06}
	&&\mathcal{J}_{\mu }^{4}(x)=g_s \overline{Q}^{a}(x)\gamma^{\theta} \frac{\lambda _{ab}^{n}}{2} G_{\mu\theta}^{n}(x)q^{b} (x)\,, \label{eq:Curr4}
\end{eqnarray}%
which couples to hybrid meson states with $0^{+(+)}$ and $1^{-(+)}$ quantum numbers. Here, $g_s$ denotes the strong coupling constant of QCD, $q=u,d,s$ and $Q=c,b$ represent the light and heavy quark fields, respectively, and $\lambda^n$ $(n=1,2,\ldots,8)$ are the Gell-Mann matrices. The dual gluon field strength tensor is defined as $\tilde{G}_{\mu\theta}^{n}(x)=\epsilon_{\mu\theta\alpha\beta}G_{\alpha\beta}^{n}(x)/2$, where $G_{\mu\theta}^{n}(x)$ denotes the gluon field strength tensor, while $a,b=1,2,3$ are color indices. 
The interpolating currents introduced above can simultaneously interpolate scalar and vector hybrid states, as well as pseudo-scalar and axial-vector hybrid states, depending on the corresponding quantum number assignments. For clarity and brevity, throughout the rest of the paper, we denote the scalar, vector, pseudo-scalar, and axial-vector hybrid mesons by $\mathrm{HM_S}$, $\mathrm{HM_V}$, $\mathrm{HM_{PS}}$, and $\mathrm{HM_{AV}}$, respectively.

Since we investigate the EM interaction of charged hybrid mesons in the presence of an external photon field, the background field approach provides a practical and reliable framework for its treatment within the light cone QCD sum rule formalism. According to the basic principles of this approach, and in the presence of a slowly varying external EM background field, $A_{\mu}(x)$, the correlation function in Eq.~(\ref{edmn01}) can be reorganized as follows:
\begin{equation}
	\label{edmn07}
	\Pi _{\mu \nu }(p,q)=i\int d^{4}x\,e^{ip\cdot x}\langle 0|\mathcal{T}\{\mathcal{J}_{\mu}(x)
	\mathcal{J}_{\nu }^{\dagger }(0)\}|0\rangle_{F}\,. 
\end{equation}%
Here, $F$ denotes the external EM background field generated by the photon, and the vacuum expectation value of the two interpolating currents is calculated in the presence of this external field. The background EM field associated with a photon of four momentum $q_\mu$ and polarization vector $\varepsilon_\mu$ can be expressed in terms of the gauge potential as,
\begin{equation}
	A_{\mu}(x)=\varepsilon_\mu e^{iq\cdot x} \,, 
\end{equation}
which gives rise to the EM field strength tensor
\begin{equation}
	F_{\mu\nu}(x)=\partial_{\mu}A_{\nu}-\partial_{\nu}A_{\mu}=-i(\varepsilon_\mu q_\nu-\varepsilon_\nu q_\mu)e^{iq\cdot x}\,.
\end{equation}
For a real photon, one has $q^2=0$ and $q\cdot\varepsilon=0$. The background field method provides a convenient and consistent framework for studying the EM interaction of hadronic systems in the presence of an external EM field. In particular, it maintains gauge invariance and allows a clean separation of the hard (perturbative) and soft (non-perturbative) photon emission contributions. The latter contributions are naturally described in terms of photon DAs of different twists, which enter the operator product expansion (OPE) and organize its long distance contributions according to the photon virtuality~\cite{Ball:2002ps}.
Treating the external EM field as a weak background, the correlation function of Eq.~(\ref{edmn07}) admits an expansion in powers of the external field, which takes the form,
\begin{equation}
	\Pi _{\mu \nu }(p,q) = \Pi _{\mu \nu }^{(0)}(p) + \Pi _{\mu \nu }^{(1)}(p,q)+... \,.
\end{equation}
The zeroth order term, $\Pi_{\mu\nu}^{(0)}$, corresponds to the correlation function in the absence of the external EM field and contains the information required for determining the masses and residues of the hybrid mesons. The linear term, $\Pi_{\mu\nu}^{(1)}$, describes the response of the correlation function to a single external photon and contains the information on the EM interaction of the hybrid mesons. Therefore, $\Pi_{\mu\nu}^{(1)}$ is the relevant contribution for extracting the MDMs and EQMs of the $\mathrm{HM_V}$ and $\mathrm{HM_{AV}}$ hybrid mesons within the light cone QCD sum rule framework.
\subsection{Hadronic representation}\label{sec:EWH}
In this section, we construct the hadronic side corresponding to Eq.~(\ref{edmn07}) for the heavy-light $\mathrm{HM_V}$ and $\mathrm{HM_{AV}}$  hybrid states.
To this end, we saturate the correlation function with complete sets of intermediate hybrid meson states having the same quantum numbers as the corresponding interpolating currents and perform the integration over $x$ in Eq.~(\ref{edmn07}). Isolating the contributions of the ground state hybrid mesons, we obtain the following expressions for the correlation functions corresponding to the different hybrid states,
\begin{align}
	\label{edmn08}
	&\Pi_{\mu\nu}^\mathrm{1,Had} (p,q) = {\frac{\langle 0 \mid \mathcal{J}_\mu (0) \mid
			\mathrm{HM_{S}}(p) \rangle}{p^2 - m_{\mathrm{HM_{S}}}^2}} \langle \mathrm{HM_{S}} (p) \mid \mathrm{HM_{S}} (p+q) \rangle_{\gamma} 
	\frac{\langle \mathrm{HM_{S}} (p+q) \mid \mathcal{J}_{\nu }^{\dagger } (0) \mid 0 \rangle}{(p+q)^2 - m_{\mathrm{HM_{S}}}^2} + \dots \\& \nonumber
	+{\frac{\langle 0 \mid \mathcal{J}_\mu (0) \mid
			\mathrm{HM_{V}}(p) \rangle}{p^2 - m_{\mathrm{HM_{V}}}^2}} \langle \mathrm{HM_{V}} (p) \mid \mathrm{HM_{V}} (p+q) \rangle_{\gamma}  
	\frac{\langle \mathrm{HM_{V}} (p+q) \mid \mathcal{J}_{\nu }^{\dagger } (0) \mid 0 \rangle}{(p+q)^2 - m_{\mathrm{HM_{V}}}^2} + \dots \,,
\end{align}
\begin{align}
	\label{edmn09}
	&\Pi_{\mu\nu}^\mathrm{2,Had} (p,q) = {\frac{\langle 0 \mid \mathcal{J}_\mu (0) \mid
			\mathrm{HM_{PS}}(p) \rangle}{p^2 - m_{\mathrm{HM_{PS}}}^2}} \langle \mathrm{HM_{PS}} (p) \mid \mathrm{HM_{PS}} (p+q) \rangle_{\gamma}  
	\frac{\langle \mathrm{HM_{PS}} (p+q) \mid \mathcal{J}_{\nu }^{\dagger } (0) \mid 0 \rangle}{(p+q)^2 - m_{\mathrm{HM_{PS}}}^2} + \dots \\& \nonumber
	+{\frac{\langle 0 \mid \mathcal{J}_\mu (0) \mid
			\mathrm{HM_{AV}}(p) \rangle}{p^2 - m_{\mathrm{HM_{AV}}}^2}} \langle \mathrm{HM_{AV}} (p) \mid \mathrm{HM_{AV}} (p+q) \rangle_{\gamma}  
	\frac{\langle \mathrm{HM_{AV}} (p+q) \mid \mathcal{J}_{\nu }^{\dagger } (0) \mid 0 \rangle}{(p+q)^2 - m_{\mathrm{HM_{AV}}}^2} + \dots \,,
\end{align}
where $m_{\mathrm{HM_{S(PS)}}}$ and $m_{\mathrm{HM_{V(AV)}}}$ denote the masses of the $\mathrm{HM_S}$ ($\mathrm{HM_{PS}}$) and $\mathrm{HM_V}$ ($\mathrm{HM_{AV}}$) hybrid states, respectively. As expected from the quantum numbers of the interpolating currents, the scalar and vector hybrid states contribute simultaneously to the corresponding hadronic representation, while the pseudo-scalar and axial-vector hybrid mesons appear together in the respective correlation function. Eqs.~(\ref{edmn08}) and~(\ref{edmn09}) take a more convenient form upon using the matrix elements given below:
\begin{align}
	\label{edmn10}
	\langle 0 | \mathcal{J}_\mu (0) | \mathrm{HM_{S(PS)}} (p) \rangle &= p_{\mu}f_{\mathrm{HM_{S(PS)}}} \,, \nonumber \\
	\langle \mathrm{HM_{S(PS)}} (p+q) | \mathcal{J}_{\nu}^{\dagger} (0) | 0 \rangle &= (p+q)_{\nu}f_{\mathrm{HM_{S(PS)}}} \,,
\end{align}%
and 
\begin{align}
	\label{edmn11}
	\langle 0|\mathcal{J}_{\mu }(0)|\mathrm{HM_{V(AV)}}(p,\varepsilon^f)\rangle &=m_{\mathrm{HM_{V(AV)}}}f_{\mathrm{HM_{V(AV)}}}\varepsilon_{\mu }^f \,, \nonumber \\
	\langle \mathrm{HM_{V(AV)}} (p+q, \varepsilon^{i}) | \mathcal{J}_{\nu}^{\dagger} (0) | 0 \rangle &= m_{\mathrm{HM_{V(AV)}}}f_{\mathrm{HM_{V(AV)}}} \, \varepsilon_\nu^{*i},
\end{align}%
where $f_{\mathrm{HM_{S(PS)}}}$ and $f_{\mathrm{HM_{V(AV)}}}$ denote the current couplings of the $\mathrm{HM_S}$ ($\mathrm{HM_{PS}}$) and $\mathrm{HM_V}$ ($\mathrm{HM_{AV}}$) hybrid mesons, respectively, while $\varepsilon_{\nu}^{i}$ ($\varepsilon_{\mu}^{f}$) represents the polarization four vector of the initial (final) $\mathrm{HM_V}$ or $\mathrm{HM_{AV}}$ hybrid state. For the respective hybrid states, the interaction with the external photon field is characterized by independent Lorentz invariant form factors, which enter the matrix elements in Eqs.~(\ref{edmn08}) and~(\ref{edmn09}) as follows~\cite{Aliev:2009gj,Azizi:2013mna,Brodsky:1992px}:
\begin{align}
	\label{edmn12}
	 \langle \mathrm{HM_{S(PS)}}(p) \mid  \mathrm{HM_{S(PS)}} (p+q)\rangle_{\gamma} &= G(Q^2)\, (2p+q)_{\tau}\, \varepsilon^\tau \,, \\ 
	 \langle \mathrm{HM_{V(AV)}}(p,\varepsilon^f) \mid  \mathrm{HM_{V(AV)}} (p+q,\varepsilon^i)\rangle_{\gamma} & = - \varepsilon^\tau (\varepsilon^{f})^\alpha (\varepsilon^{i})^\beta
	 \Big[ G_1(Q^2) (2p+q)_\tau \, g_{\alpha\beta}
	 + G_2(Q^2) ( g_{\tau\beta}\, q_\alpha - g_{\tau\alpha}\, q_\beta ) \nonumber\\
	 &\quad - \frac{1}{2 m_{\mathrm{HM_{V(AV)}}}^2} G_3(Q^2) (2p+q)_\tau \, q_\alpha q_\beta \Big]\,.
	\label{edmn13}
\end{align}
Here, $\varepsilon^\tau$ denotes the polarization four vector of the photon, while $Q^2=-q^2$ represents the squared momentum transfer carried by the photon.

As an illustrative example, we first derive the hadronic representation of the correlation function for the $\mathrm{HM_S}$ and $\mathrm{HM_V}$ hybrid states and then generalize the procedure to the $\mathrm{HM_{PS}}$ and $\mathrm{HM_{AV}}$ states. To this end, we substitute the scalar and vector parts of Eqs.~(\ref{edmn10})--(\ref{edmn13}) into Eq.~(\ref{edmn08}), which represents the hadronic correlation function for the scalar and vector hybrid mesons. This procedure leads to,
\begin{align}
	\label{edmn14}
	&\Pi_{\mu\nu}^\mathrm{1,Had}(p,q) = \frac{f_{\mathrm{HM_{S}}}^2 \, \varepsilon^\tau}{ [m_{\mathrm{HM_{S}}}^2 - (p+q)^2][m_{\mathrm{HM_{S}}}^2 - p^2]}\, G(Q^2)\, p_\mu\, (p+q)_\nu\, (2p+q)_\tau + \dots \nonumber \\&
	+  \frac{m_{\mathrm{HM_{V}}}^2 f_{\mathrm{HM_{V}}}^2 \, \varepsilon^\tau}{ [m_{\mathrm{HM_{V}}}^2 - (p+q)^2][m_{\mathrm{HM_{V}}}^2 - p^2]}
	\bigg\{G_1(Q^2)(2p+q)_\tau\bigg[-g_{\mu\nu}+\frac{p_\mu p_\nu}{m_{\mathrm{HM_{V}}}^2}
	+\frac{(p+q)_\mu (p+q)_\nu}{m_{\mathrm{HM_{V}}}^2}-\frac{p_\mu(p+q)_\nu}{2m_{\mathrm{HM_{V}}}^4}\nonumber\\
	& (Q^2+2m_{\mathrm{HM_{V}}}^2)
	\bigg]
	+ G_2 (Q^2) \bigg[-q_\mu g_{\tau\nu}  
	+ q_\nu g_{\tau\mu}-
	\frac{p_\mu}{m_{\mathrm{HM_{V}}}^2}  \big(q_\nu p_\tau - \frac{1}{2}
	Q^2 g_{\nu\tau}\big) 
	+
	\frac{(p+q)_\nu}{m_{\mathrm{HM_{V}}}^2}  \big(q_\mu (p+q)_\tau+ \frac{1}{2}
	Q^2 g_{\mu\tau}\big) 
	\nonumber\\
	&-  
	\frac{p_\mu(p+q)_\nu p_\tau}{m_{\mathrm{HM_{V}}}^4} \, Q^2
	\bigg]
	+\frac{G_3(Q^2)}{2m_{\mathrm{HM_{V}}}^2}(2p+q)_\tau \bigg[
	q_\mu q_\nu -\frac{p_\mu q_\nu}{2 m_{\mathrm{HM_{V}}}^2} Q^2 
	+\frac{q_\mu (p+q)_\nu}{2 m_{\mathrm{HM_{V}}}^2} Q^2
	-\frac{p_\mu(p+q)_\nu}{4 m_{\mathrm{HM_{V}}}^4} Q^4\bigg]
	\bigg\} + \dots\,.
\end{align}
As can be seen, both $\mathrm{HM_S}$ and $\mathrm{HM_V}$ hybrid mesons contribute to the hadronic representation of the correlation function. To isolate the scalar contribution, we project the correlation function in Eq.~(\ref{edmn14}) by contracting it with $p^\mu(p+q)^\nu$. This projection eliminates the vector hybrid meson contribution and yields,
\begin{align}
\label{edmn14a}
&p^\mu(p+q)^\nu\ \Pi_{\mu\nu}^\mathrm{1,Had}(p,q) = \frac{f_{\mathrm{HM_{S}}}^2 \, m_{\mathrm{HM_{S}}}^4\, \varepsilon^\tau}{ [m_{\mathrm{HM_{S}}}^2 - (p+q)^2][m_{\mathrm{HM_{S}}}^2 - p^2]}\, G(Q^2)\, (2p+q)_\tau + \dots\,.
\end{align}
The hadronic representation of the vector hybrid states can be isolated by selecting the Lorentz structures in Eq.~(\ref{edmn14}) that are free from contributions of the scalar hybrid states and receive contributions solely from the vector hybrid states, which yields:
\begin{align}
\label{edmn14b}
&\Pi_{\mu\nu}^\mathrm{V,Had}(p,q) = \frac{m_{\mathrm{HM_{V}}}^2 f_{\mathrm{HM_{V}}}^2 \, \varepsilon^\tau}{ [m_{\mathrm{HM_{V}}}^2 - (p+q)^2][m_{\mathrm{HM_{V}}}^2 - p^2]}
\bigg\{G_1(Q^2)(2p+q)_\tau\bigg[-g_{\mu\nu}+\frac{p_\mu p_\nu}{m_{\mathrm{HM_{V}}}^2}
+\frac{q_\mu (p+q)_\nu}{m_{\mathrm{HM_{V}}}^2}\bigg]
\nonumber\\& 
+ G_2 (Q^2) \bigg[-q_\mu g_{\tau\nu}  
+ q_\nu g_{\tau\mu}-
\frac{p_\mu}{m_{\mathrm{HM_{V}}}^2}  \big(q_\nu p_\tau - \frac{1}{2}
Q^2 g_{\nu\tau}\big) 
+
\frac{(p+q)_\nu}{m_{\mathrm{HM_{V}}}^2}  \big(q_\mu (p+q)_\tau+ \frac{1}{2}
Q^2 g_{\mu\tau}\big) 
\bigg]\nonumber\\
&
+\frac{G_3(Q^2)}{2m_{\mathrm{HM_{V}}}^2}(2p+q)_\tau \bigg[
q_\mu q_\nu -\frac{p_\mu q_\nu}{2 m_{\mathrm{HM_{V}}}^2} Q^2 
+\frac{q_\mu (p+q)_\nu}{2 m_{\mathrm{HM_{V}}}^2} Q^2 \bigg] 
\bigg\} + \dots\,.
\end{align}

From this point onward, our discussion and calculations in the hadronic section are based on the obtained representation of the vector hybrid mesons in Eq.~(\ref{edmn14b}). Since the primary aim of this work is to examine the EM properties of the hybrid states, it is useful to rewrite the form factors $G_i(Q^2)$ appearing in Eq.~(\ref{edmn14b}) in terms of the magnetic and quadrupole form factors, $F_M(Q^2)$ and $F_{\mathcal{D}}(Q^2)$, in the following form~\cite{Brodsky:1992px}:
\begin{align}
	\label{edmn15}
	F_M(Q^2) &= G_2(Q^2), \nonumber \\
	F_{\mathcal{D}}(Q^2) &= G_1(Q^2) - G_2(Q^2) + \left(1 + \frac{Q^2}{4m_{\mathrm{{HM}_{V}}}^2}\right) G_3(Q^2).
\end{align}
In the static limit, $Q^2=0$, and after substituting Eq.~(\ref{edmn15}) into Eq.~(\ref{edmn14b}) and retaining the Lorentz structures containing $F_M(0)$ and $F_{\mathcal{D}}(0)$, the correlation function for the vector hybrid mesons takes the following form on the hadronic side,
\begin{align}
	\label{edmn16}
	\Pi_{\mu\nu}^\mathrm{V,Had}(p,q) = \frac{m_{\mathrm{HM_{V}}}^2 f_{\mathrm{HM_{V}}}^2} {\big[m^{2}_{\mathrm{HM_{V}}} - (p+q)^2\big]\big[m^{2}_{\mathrm{HM_{V}}} - p^{2}\big]}
	\Big[ F_{M}(0)\Big( q_{\nu}\varepsilon_{\mu} - q_{\mu}\varepsilon_{\nu} + \frac{p\cdot \varepsilon}{m^{2}_{\mathrm{HM_{V}}}} (p_{\nu}q_{\mu} &- p_{\mu}q_{\nu}) \Big) \nonumber \\
	+ F_{\mathcal{D}}(0) \frac{p\cdot \varepsilon}{m^{2}_{\mathrm{HM_{V}}}} \, q_{\mu}q_{\nu} \Big] + \dots\,.
\end{align}
The unwanted contributions from higher resonances and the continuum in the correlation function above are suppressed by performing the double Borel transformation defined as,
\begin{equation}\label{BorelTHRep}
\mathcal{B}_{p_1^2}(M_1^2) \mathcal{B}_{p_2^2}(M_2^2) \frac{1}{(p_1^2-m_1^2)^n}\frac{1}{(p_2^2-m_2^2)^m}\rightarrow (-1)^{n+m} \frac{1}{\Gamma[n]\Gamma[m]} \frac{1}{(M_1^2)^{n-1}} \frac{1}{(M_2^2)^{m-1}} e^{-\frac{m_1^2}{M_1^2}} e^{-\frac{m_2^2}{M_2^2}} \,.
\end{equation}
Applying this double Borel transformation with respect to $p^2$ and $(p+q)^2$ to the correlation function in Eq.~(\ref{edmn16}), we obtain,
\begin{align}\label{edmn16a}
\mathcal{B}_{p^2}(M_1^2) \mathcal{B}_{(p+q)^2}(M_2^2)\,\Pi_{\mu\nu}^\mathrm{V,Had}(p,q) =\, & m_{\mathrm{HM_{V}}}^2 f_{\mathrm{HM_{V}}}^2 e^{-\frac{m_{\mathrm{HM_{V}}}^2}{M_1^2}} e^{-\frac{m_{\mathrm{HM_{V}}}^2}{M_2^2}}
\Big[ F_{M}(0)\Big( q_{\nu}\varepsilon_{\mu} - q_{\mu}\varepsilon_{\nu} + \frac{p\cdot \varepsilon}{m^{2}_{\mathrm{HM_{V}}}} \nonumber \\& (p_{\nu}q_{\mu} - p_{\mu}q_{\nu}) \Big) 
+ F_{\mathcal{D}}(0) \frac{p\cdot \varepsilon}{m^{2}_{\mathrm{HM_{V}}}} \, q_{\mu}q_{\nu} \Big] + \dots\,,
\end{align}
where $M_1^2$ and $M_2^2$ are the Borel mass parameters, and the ellipsis denotes the contributions from higher resonances and the hadronic continuum. 

The MDMs and EQMs are then defined through the static values of the corresponding EM form factors at $Q^2=0$:
\begin{align}\label{edmn17}
	\mu_{} &= \frac{e}{2\, m_{\mathrm{{HM}_{V}}}}\, F_M(0), \\
	\mathcal{D}_{} &= \frac{e}{m_{\mathrm{{HM}_{V}}}^2}\, F_{\mathcal{D}}(0).
	\label{edmn17a}
\end{align}
The same procedure can be followed for the $\mathrm{HM_{PS}}$ and $\mathrm{HM_{AV}}$ hybrid mesons. The resulting expressions are analogous to Eqs.~(\ref{edmn14})--(\ref{edmn17a}), with the corresponding hadronic parameters $m_{\mathrm{HM_{PS(AV)}}}$ and $f_{\mathrm{HM_{PS(AV)}}}$ replacing those of the scalar and vector states. The remaining step in constructing the hadronic representation is the treatment of the higher resonance and continuum contributions through the double Borel transformation and continuum subtraction. We postpone the detailed discussion of this procedure to section~\ref{sec24}.
\subsection{QCD representation}\label{sec:QCDF}
The initial QCD representation of the correlation function is evaluated in the deep Euclidean region within the OPE framework. For this purpose, the interpolating currents associated with the hybrid mesons of different quantum numbers, specified in Eqs.~(\ref{edmn03})--(\ref{edmn06}) are inserted into Eq.~(\ref{edmn07}), after which the quark field operators are fully contracted according to Wick's theorem. For the different hybrid states considered in this study, the resulting correlation functions are given by,
\begin{align}
	\Pi _{\mu \nu }^{1,\mathrm{OPE}}(p,q)=i\frac{\epsilon_{\mu\theta\alpha\beta}\epsilon_{\nu\delta\alpha^{\prime }\beta^{\prime }}}{4} & \int d^{4}x\, e^{ipx}\, \frac{\lambda^{n}_{ab} \, \lambda^{m}_{a^{\prime }b^{\prime}}}{4} \, 
	 \langle 0|g_s^{2} G^n_{\alpha \beta}(x) G^m_{\alpha^{\prime }\beta^{\prime }}(0)|0 \rangle \nonumber \\& 
	\times \langle 0| \mathrm{Tr}\left[\gamma^{\theta }\gamma _{5}S_{q}^{bb^{\prime }}(x) \gamma_5 \gamma^{\delta} S_{Q}^{a^{\prime }a}(-x) \right]|0 \rangle_\gamma ,  
    \label{eq:OPE1}
\end{align}%
\begin{align}
	&\Pi _{\mu \nu }^{2,\mathrm{OPE}}(p,q)=i\int d^{4}x\, e^{ipx}\frac{\lambda^{n}_{ab} \, \lambda^{m}_{a^{\prime }b^{\prime}}}{4} \, \langle 0|g_s^{2} G^n_{\mu \theta}(x) G^m_{\nu \delta}(0)|0 \rangle \, \langle 0| \mathrm{Tr}\left[\gamma^{\theta}\gamma_{5}S_{q}^{bb^{\prime}}(x)\gamma_{5}\gamma^{\delta} S_{Q}^{a^{\prime }a}(-x)\right]|0\rangle_\gamma \,, 
	  \label{eq:OPE2}
\end{align}%
\begin{align}
	\Pi _{\mu \nu }^{3,\mathrm{OPE}}(p,q)=-i\frac{\epsilon_{\mu\theta\alpha\beta}\epsilon_{\nu\delta\alpha^{\prime }\beta^{\prime }}}{4} & \int d^{4}x\, e^{ipx}\, \frac{\lambda^{n}_{ab} \, \lambda^{m}_{a^{\prime }b^{\prime}}}{4} \, \langle 0|g_s^{2}G^n_{\alpha \beta}(x) G^m_{\alpha^{\prime }\beta^{\prime }}(0)|0 \rangle\, \nonumber \\&
	\times \langle 0| \mathrm{Tr}\left[\gamma^{\theta }S_{q}^{bb^{\prime}}(x)\gamma^{\delta} S_{Q}^{a^{\prime }a}(-x)\right])|0 \rangle_\gamma,  
     \label{eq:OPE3}
\end{align}%
and
\begin{align}
	&\Pi _{\mu \nu }^{4,\mathrm{OPE}}(p,q)=-i\int d^{4}x\, e^{ipx}\, \frac{\lambda^{n}_{ab} \, \lambda^{m}_{a^{\prime }b^{\prime}}}{4} \, \langle 0|g_s^{2}G^n_{\mu \theta}(x) G^m_{\nu \delta}(0)|0 \rangle\, \langle 0| \mathrm{Tr}\left[\gamma^{\theta}S_{q}^{bb^{\prime}}(x)\gamma^{\delta} S_{Q}^{a^{\prime}a}(-x)\right] |0 \rangle_\gamma,  
	  \label{eq:OPE4}
\end{align}%
where $S_{q(Q)}(x)$ denotes the light (heavy) quark propagator. It should also be noted that Eqs.~(\ref{eq:OPE1})--(\ref{eq:OPE4}) correspond, respectively, to the interpolating currents $\mathcal J_\mu^1$, $\mathcal J_\mu^2$, $\mathcal J_\mu^3$, and $\mathcal J_\mu^4$ given in Eqs.~(\ref{edmn03})--(\ref{edmn06}). In coordinate space, the light and heavy quark propagators can be expressed as follows~\cite{Azizi:2018mte,Ozdem:2024dbq,Ozdem:2025ncd}:
\begin{align}
	\label{edmn18}
	S_{q}(x)&= S_q^{free}(x) 
	- \frac{\langle \bar qq \rangle }{12} \Big(1-i\frac{m_{q} \slashed{x}}{4}   \Big)
	- \frac{ \langle \bar qq \rangle }{192}
	m_0^2 x^2  \Big(1 
	-i\frac{m_{q} \slashed{x}}{6}   \Big)
	+\frac {i g_s~G^{\mu \nu} (x)}{32 \pi^2 x^2} 
	\Bigg[\rlap/{x} 
	\sigma_{\mu \nu} +  \sigma_{\mu \nu} \rlap/{x}
	\Bigg],\\
	%
	S_{Q}(x)&=S_Q^{free}(x)
	-i \frac{m_{Q}\,g_{s}}{16\pi ^{2}} \int_{0}^{1} d\upsilon\; G^{\mu\nu}(\upsilon x)
	\Bigg[ (\sigma_{\mu\nu}\rlap/{x} + \rlap/{x}\sigma_{\mu\nu})
	\frac{K_{1}\!\big(m_{Q}\sqrt{-x^{2}}\big)}{\sqrt{-x^{2}}}
	+ 2\sigma_{\mu\nu} K_{0}\!\big(m_{Q}\sqrt{-x^{2}}\big) \Bigg],
	\label{edmn19}
\end{align}
with  
\begin{align}
	S_q^{free}(x)&=\frac{1}{2 \pi^2 x^2}\Big(i \frac{\slashed{x}}{x^2}- \frac{m_q}{2}\Big),\\
	\nonumber\\
	S_Q^{free}(x)&=\frac{m_{Q}^{2}}{4 \pi^{2}} \Bigg[ \frac{K_{1}\big(m_{Q}\sqrt{-x^{2}}\big) }{\sqrt{-x^{2}}}
	+i\frac{{\slashed{x}}~K_{2}\big( m_{Q}\sqrt{-x^{2}}\big)}
	{(\sqrt{-x^{2}})^{2}}\Bigg],
\end{align}
where the gluon field strength tensor is denoted by $G_{\mu\nu}$, while $K_n(z)$ represents the modified Bessel function of the second kind.

The correlation functions in Eqs.~(\ref{eq:OPE1})--(\ref{eq:OPE4}), which describe the interaction of the external EM background field with the quark content of the hybrid mesons, receive contributions from two distinct photon emission mechanisms on the QCD side. In the first mechanism, the photon interacts perturbatively with a light or heavy quark at short distances. This contribution is referred to as the perturbative photon emission contribution. In this case, one of the free light or heavy quark propagators appearing in Eqs.~(\ref{eq:OPE1})--(\ref{eq:OPE4}) is replaced by the corresponding propagator with a photon insertion, given by,
\begin{align}
	\label{free}
	S^{free}(x) \longrightarrow \int d^4y\, S^{free} (x-y)\,\rlap/{\!A}(y)\, S^{free} (y)\,,
\end{align} 
while the other quark propagator is taken in its full form. Substituting these expressions into the correlation functions and carrying out the relevant traces and calculations yields the perturbative contribution associated with the short distance interaction of the photon with a quark.

The second mechanism corresponds to the interaction of the photon with the quark fields at long distances and therefore gives rise to non-perturbative contributions. To account for these contributions, the light quark propagator is replaced by,
\begin{align}
	\label{edmn21}
	S_{\alpha\beta}^{ab}(x) \longrightarrow -\frac{1}{4} \big[\bar{q}^a(x) \Gamma_i q^b(0)\big]\big(\Gamma_i\big)_{\alpha\beta}\,,
\end{align}
where $\Gamma_i={\mathbf{1},\gamma_5,\gamma_\mu,i\gamma_5\gamma_\mu,\sigma_{\mu\nu}/2}$. The other quark propagator appearing in Eqs.~(\ref{eq:OPE1})--(\ref{eq:OPE4}) is retained in its full form. After making these replacements and performing the relevant traces and calculations, the non-perturbative photon emission contributions to the QCD representation can be obtained. Several points concerning the non-perturbative contribution are worth noting. After carrying out the corresponding calculations, matrix elements of the forms $\langle \gamma(q)| \bar{q}(x) \Gamma_i G_{\alpha\beta}q(0) | 0\rangle$ and $\langle \gamma(q)| \bar{q}(x) \Gamma_i q(0) | 0\rangle$ arise in the correlation functions. These non-local matrix elements describe the long distance interaction of the on-shell photon with the quark and gluon fields and are parameterized in terms of the photon DAs of different twists. The explicit parametrizations of these matrix elements in terms of the corresponding photon DAs are collected in Appendix~\ref{AppenPDAs}; further details can be found in Refs.~\cite{Ball:2002ps,Rohrwild:2007iz}. Finally, owing to the large masses of the heavy quarks and the resulting suppression of their long distance photon emission, the non-perturbative photon contributions associated with the heavy quark propagators are neglected in our analysis.

In evaluating the perturbative and non-perturbative photon emission contributions to the QCD representation, the gluon field matrix element $ \langle 0 |G^n_{\alpha \beta}(x)G^m_{\alpha' \beta'}(0)|0 \rangle $ appearing in Eqs.~(\ref{eq:OPE1})--(\ref{eq:OPE4}) is considered through two types of contributions. First, it is replaced by the free gluon propagator in coordinate space,
\begin{align}\label{eq:Gprop}
	\langle 0 |G^n_{\alpha \beta}(x)G^m_{\alpha' \beta'}(0)||0 \rangle =
	\frac{\delta^{mn}}{2 \pi^2 x^4} \big[g_{\beta \beta'}(g_{\alpha \alpha'}-\frac{4 x_{\alpha} x_{\alpha'}}{x^2}) +(\beta, \beta') \leftrightarrow (\alpha, \alpha')
	-\beta \leftrightarrow \alpha -\beta' \leftrightarrow \alpha' \big]\,, 
\end{align} 
which accounts for the perturbative contribution. Second, the gluon field matrix element is approximated by its leading term in the Taylor expansion around $x=0$, which is expressed in terms of the gluon condensate as,
\begin{align}
	\label{eq:Gcond}
	&&\langle 0 |G^n_{\alpha \beta}(0)G^m_{\alpha' \beta'}(0)|0 \rangle =
	\frac{\langle G^2\rangle }{96}\delta^{mn} [g_{\alpha \alpha'} g_{\beta \beta'}-g_{\alpha \beta'} g_{\alpha'\beta }]\,,
\end{align}
which accounts for the non-perturbative contribution. In addition, the following color algebra relation is used throughout the calculations,
\begin{align}
	\label{eq:tntn}
	&&\mathrm t^{n}_{ab} \mathrm t^{n}_{a^{\prime }b^{\prime }}=\frac{\lambda^{n}_{ab} \, \lambda^{n}_{a^{\prime }b^{\prime}}}{4}=\frac{1}{2}\left(\delta^{ab'}\delta^{a'b}-\frac{1}{3}\delta^{ab}\delta^{a'b'}\right)\,.
\end{align}

As an illustrative example, after performing the required calculations and evaluating the perturbative and non-perturbative photon emission contributions for the current $\mathcal J_\mu^1$ in Eq.~(\ref{edmn03}) and the corresponding correlation function in Eq.~(\ref{eq:OPE1}), the QCD representation takes the form,
\begin{align}
\Pi _{\mu \nu }^{1,\mathrm{OPE}}(p,q)= \Pi _{\mu \nu }^{1,\mathrm{pert}}(s_1,s_2) + \Pi _{\mu \nu }^{1,\mathrm{nonpert}}(s_1,s_2) = \Pi _{\mu \nu }^{1,\mathrm{QCD}}(s_1,s_2) \,. 
\label{eq:CFQCD1}
\end{align}%
The perturbative and non-perturbative contributions can then be represented through the corresponding spectral densities by means of a double dispersion relation. Applying the double Borel transformation with respect to $p^2$ and $(p+q)^2$, using the basic relation,
\begin{equation}
\label{BorelTQRep}
\mathcal{B}_{P^{2}}(M^2)\, e^{-\alpha P^{2}} = \delta(\frac{1}{M^2}-\alpha) \,,
\end{equation}
and performing the required mathematical manipulations, we obtain:
\begin{align}
\label{eq:CFQCD1a}
\mathcal{B}_{p^2}(M_1^2) \mathcal{B}_{(p+q)^2}(M_2^2)\,\Pi _{\mu \nu }^{1,\mathrm{OPE}}(p,q) =  & \int_0^\infty ds_1 \int_0^\infty ds_2 \, e^{-\frac{s_1}{M_1^2}-\frac{s_2}{M_2^2}}\rho^{1,\mathrm{QCD}}(s_1,s_2) 
\,.
\end{align}
Here, $s_1$ and $s_2$ are the spectral variables associated with the initial and final hybrid states, respectively, while $\rho^{1,\mathrm{QCD}}(s_1,s_2)$ denotes the total QCD spectral density, containing both the perturbative and non-perturbative photon emission contributions. The continuum thresholds associated with the initial and final states will be denoted separately by the corresponding threshold parameters and will be introduced when the continuum subtraction is performed.

\begin{figure}[h!]
	\centering
	\includegraphics[width=0.85\textwidth]{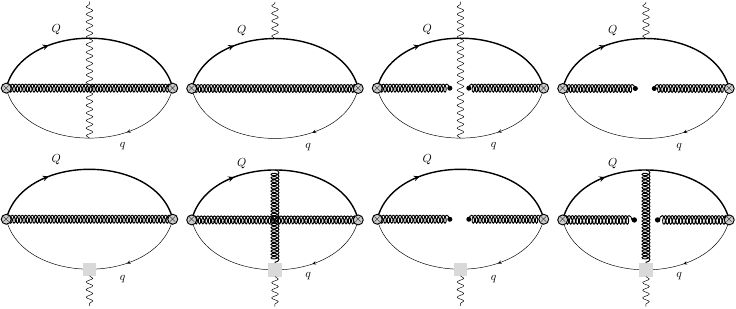}
	\caption{Representative Feynman diagrams contributing to the perturbative (first row) and non-perturbative (second row) photon emission contributions considered in this work.}
	\label{FeynDiag}
\end{figure}

The same procedure is applied to the remaining correlation functions in Eqs.~(\ref{eq:OPE1})--(\ref{eq:OPE4}), yielding the perturbative and non-perturbative photon emission contributions for all hybrid configurations considered in this work. Some representative Feynman diagrams contributing to these calculations are depicted in Figure~\ref{FeynDiag}. With the QCD representations thus established, they can be matched to the corresponding hadronic representations to derive the light cone QCD sum rules for the magnetic dipole and electric quadrupole form factors. The details of the matching procedure and continuum subtraction are presented in the following section~\ref{sec24}.
\subsection{Light cone sum rules}\label{sec24}
The light cone QCD sum rules for the magnetic dipole and electric quadrupole form factors are obtained by considering the Lorentz structures $q_{\mu}\varepsilon_{\nu}$ and $(\varepsilon\cdot p)q_{\mu}q_{\nu}$ in both the QCD and hadronic representations, respectively. For the magnetic form factor, we first equate the coefficients of the Lorentz structure $q_{\mu}\varepsilon_{\nu}$ on both sides, which leads to,
\begin{align}
-m_{i}^2 f_{i}^2 \, F_M(0)\, e^{-\frac{m_{i}^2}{M_1^2}} e^{-\frac{m_{i}^2}{M_2^2}} + \dots &= \int_0^\infty ds_1 \int_0^\infty ds_2 \, e^{-\frac{s_1}{M_1^2}-\frac{s_2}{M_2^2}}\rho_i^{QCD}(s_1,s_2)\,,
\label{correlated}
\end{align}
where the index $i$ specifies the corresponding hybrid state and its associated QCD representation.

To isolate the ground state contribution, the effects of higher resonances and the hadronic continuum are treated within the quark-hadron duality approximation. Within this prescription, the physical double spectral density is approximated by the corresponding QCD spectral density above the appropriate continuum thresholds,
\begin{equation}\label{contin1}
\rho^{\rm phys}(s_1,s_2) \simeq \rho^{\rm QCD}(s_1,s_2)\,.
\end{equation}
Since the correlation function depends on two independent dispersion variables, the continuum contribution is, in principle, defined in the two dimensional $(s_1,s_2)$ plane. Instead of introducing an arbitrary rectangular duality region, we formulate the continuum subtraction in terms of the effective spectral variable naturally associated with the double Borel transformation. We define,
\begin{equation}
s=u_0s_1+\bar u_0s_2\,, \qquad u_0=\frac{M_2^2}{M_1^2+M_2^2}\,, \qquad \bar u_0 = 1 - u_0 \,,
\label{contin2}
\end{equation}
and,
\begin{equation}\label{contin3} 
M^2 = \frac{M_1^2 M_2^2}{M_1^2 + M_2^2}\,.
\end{equation}
With these definitions, the Borel exponential factor can be written as,
\begin{equation}%
	\label{cont4}
\frac{s_1}{M_1^2} + \frac{s_2}{M_2^2} = \frac{u_0 s_1 + \bar u_0 s_2}{M^2} = \frac{s}{M^2}\,.
\end{equation}
Thus, the exponential factor in the double dispersion representation depends only on the effective variable $s$. To make this reduction explicit, we introduce the auxiliary variable,
\begin{equation}\label{contin5}
u = \frac{u_0 s_1}{s}\,,
\end{equation}
which gives,
\begin{equation} 
s_1 = \frac{su}{u_0}\,, \qquad s_2=\frac{s(1-u)}{\bar u_0}\,,
\label{contin6}
\end{equation}
with the corresponding Jacobian,
\begin{equation}
	\label{contin7}
ds_1 ds_2 = \frac{s}{u_0 \bar u_0}\, ds\, du\,.
\end{equation}
Consequently, the double dispersion integral can be recast in an effective single dispersion form,
\begin{align}\label{contin8} 
&\int_0^\infty ds_1\int_0^\infty ds_2\, e^{-\frac{s_1}{M_1^2}-\frac{s_2}{M_2^2}} \rho^{\rm QCD}(s_1,s_2) = \int_0^\infty ds\, e^{-\frac{s}{M^2}}\, \rho^{\rm eff,QCD}(s)\,,
\end{align}
where the effective spectral density is defined as,
\begin{align}
	\label{contin9}
\rho^{\rm eff,QCD}(s) = \frac{s}{u_0\bar u_0} \int_0^1 du\, \rho^{\rm QCD} \left(\frac{su}{u_0}\,, \frac{s(1-u)}{\bar u_0} \right)\,.
\end{align}
The continuum subtraction can then be implemented by restricting the effective spectral variable to the region below the continuum threshold $s_0$:
\begin{align}
\int_0^\infty ds\, \rho^{\rm eff,QCD}(s)\, e^{-s/M^2} \longrightarrow \int_{s_{\rm min}}^{s_0}ds\, \rho^{\rm eff,QCD}(s)\, e^{-s/M^2}\,.
\label{contin10}
\end{align}
In terms of the original double spectral variables, this prescription corresponds to retaining the triangular duality region,
\begin{equation}
u_0 s_1 + \bar u_0 s_2 < s_0\,,
\label{contin11}
\end{equation}
while the complementary region is identified with the higher resonances and hadronic continuum and is subtracted according to the quark-hadron duality approximation.

Since the initial and final hybrid states considered in this work have equal masses, we adopt the choice,
\begin{equation}
	\label{contin12}
M_1^2 = M_2^2 = 2 M^2\,,
\end{equation}
which yields,
\begin{equation}
u_0 = \bar u_0 = \frac12\,, \qquad s = \frac{s_1 + s_2}{2}\,.
\label{contin13}
\end{equation}
Accordingly, the continuum boundary in the original $(s_1,s_2)$ plane is given by $s_1 + s_2 = 2s_0$.

Applying the above continuum subtraction prescription to Eq.~(\ref{correlated}), the light cone QCD sum rule for the magnetic form factor takes the form,
\begin{align}
-\, m_{i}^2 f_{i}^2 \, F_M(0)\, e^{-\frac{m_{i}^2}{M^2}} &= \Pi_i^{\rm QCD}(\rm{M^2},\rm{s_0})\,.
\label{correlateda}
\end{align}
Using Eq.~(\ref{edmn17}), the corresponding light cone QCD sum rule for the MDM is obtained as,
\begin{align}
 \mu_i  &= -\frac{e^{\frac{m_{i}^2}{M^2}}}{m_{i}^2 f_{i}^2}\, \Pi_i^{\rm QCD}(\rm{M^2},\rm{s_0})\,.
\label{LCSRMM}
\end{align}
Following the same procedure for the electric quadrupole form factor and considering the Lorentz structure $(\varepsilon\cdot p)q_\mu q_\nu$, we obtain the corresponding light cone sum rule for the EQM,
\begin{align}
\mathcal{D}_i  &= \frac{e^{\frac{m_{i}^2}{M^2}}}{f_{i}^2}\, \Pi_i^{\prime \rm QCD}(\rm{M^2},\rm{s_0})\,.
\label{LCSRQM}
\end{align}
The explicit expressions for $\Pi_i^{\rm QCD}(\rm{M^2},\rm{s_0})$ and $\Pi_i^{\prime\rm QCD}(\rm{M^2},\rm{s_0})$ corresponding to the current $\mathcal J_\mu^1$ in Eq.~(\ref{edmn03}) are presented in Appendix~\ref{AppenCorQCD}. The corresponding expressions for the remaining interpolating currents are obtained following the same procedure. Further details concerning the double Borel transformation and continuum subtraction procedure can be found in Refs.~\cite{Agaev:2016srl,Ozdem:2017jqh,Azizi:2018duk,Ozdem:2024dbq}.
\section{Numerical Results}\label{sec:Num}
In this section, we first provide the values of the input parameters adopted in the numerical analysis. Using these parameters, we then perform a detailed analysis of the pole contribution and OPE convergence, and subsequently determine the MDMs and EQMs of the different vector and axial-vector hybrid meson configurations.
\subsection{Input parameters}
As mentioned in the Introduction, two main studies~\cite{Ho:2016owu,Barsbay:2022gtu} have investigated possible heavy-light hybrid meson configurations using QCD sum rule approaches. In Ref.~\cite{Ho:2016owu}, the QCD Laplace sum rule method was employed, whereas Ref.~\cite{Barsbay:2022gtu} utilized the QCD Borel sum rule approach. In both studies, a systematic analysis of heavy-light hybrid mesons was performed, and their corresponding physical parameters, including the masses and current couplings, were determined. In the present work, we adopt the results of Ref.~\cite{Barsbay:2022gtu} as input parameters for our numerical analysis. In that study, the QCD Borel sum rules were analyzed by including higher dimensional condensate contributions up to dimension ten and carefully determining the working regions of the auxiliary parameters. The authors found that all the heavy-light hybrid configurations considered in their analysis exhibit stable sum rule predictions, in contrast to the findings of Ref.~\cite{Ho:2016owu}, where some of the hybrid configurations were found to be unstable. The masses and current couplings of the heavy-light $\mathrm{HM_V}$ and $\mathrm{HM_{AV}}$ hybrid mesons with different quantum numbers, as obtained in Ref.~\cite{Barsbay:2022gtu}, are summarized in Tables~\ref{C0_results_table}--\ref{B0_results_table}. The remaining input parameters employed in the numerical analysis are collected in Table~\ref{InParam}, while the parameters entering the photon DAs employed in the calculations are presented in Appendix~\ref{AppenPDAs}.
\begin{table}[tbp]
	\centering
	\caption{Masses and current couplings of the ground state vector and axial-vector charm-nonstrange hybrid mesons with different quantum numbers.}
	\label{C0_results_table}
	\begin{tabular}{|c|c|c|c|c|} \hline\hline
		$J^{P(C)}$ & $1^{+(+)}$ & $1^{+(-)}$  & $1^{-(-)}$ & $1^{-(+)}$\\  \hline \hline
		$m_i \pm \delta m_{i}\ (\rm GeV)$ & $4.22\pm 0.15$ & $3.88\pm 0.23$  & $4.36\pm 0.15$ & $ 3.91\pm 0.22$\\
		$(f_{i}\pm \delta f_{i})\times10\ (\rm GeV^3)$ & $ 0.64\pm 0.13$  & $0.73\pm 0.14$  & $ 0.65\pm 0.14$ & $ 0.74\pm 0.13$ \\
		\hline \hline
	\end{tabular}
\end{table}
\begin{table}[tbp]
	\centering
	\caption{Masses and current couplings of the ground state vector and axial-vector charm-strange hybrid mesons with different quantum numbers. }
	\label{Cs_results_table}
	\begin{tabular}{|c|c|c|c|c|} \hline\hline
		$J^{P(C)}$ & $1^{+(+)}$ & $1^{+(-)}$  & $1^{-(-)}$ & $1^{-(+)}$\\  \hline \hline
		$m_i \pm \delta m_{i}\ (\rm GeV)$ & $4.38\pm 0.13$ & $4.05\pm 0.23$  & $4.44\pm 0.13$ & $ 4.03\pm 0.22$\\
		$(f_{i}\pm \delta f_{i})\times10\ (\rm GeV^3)$ & $ 0.77\pm 0.14$  & $0.83\pm 0.15$  & $ 0.76\pm 0.13$ & $ 0.86\pm 0.15$ \\
		\hline\hline
	\end{tabular}
\end{table}
\begin{table}[tbp]
	\centering
	\caption{Masses and current couplings of the ground state vector and axial-vector bottom-nonstrange hybrid mesons with different quantum numbers.}
	\label{B0_results_table}
	\begin{tabular}{|c|c|c|c|c|}\hline\hline
		$J^{P(C)}$ & $1^{+(+)}$ & $1^{+(-)}$  & $1^{-(-)}$ & $1^{-(+)}$\\  \hline \hline
		$m_i \pm \delta m_{i}\ (\rm GeV)$ & $8.24\pm 0.30$ & $8.08\pm 0.35$  & $8.32\pm 0.26$ & $ 8.07\pm 0.34$\\
		$(f_{i}\pm \delta f_{i})\times10\ (\rm GeV^3)$ & $ 2.65\pm 0.57$  & $2.56\pm 0.61$  & $ 2.70\pm 0.54$ & $ 2.58\pm 0.35$ \\
		\hline\hline
	\end{tabular}
\end{table}
\begin{table}[tbp]
\centering
\caption{Values of the remaining input parameters employed in the numerical analysis.}
\label{InParam}
\begin{tabular}{|c|c|} \hline \hline
	\text{Parameters} & \text{Values}  \\ \hline \hline
	$m_u=m_d$    & $0$ \\
	$m_s$    & $(93.4_{-3.4}^{+8.6})\ $MeV \cite{ParticleDataGroup:2024cfk}     \\ 
	$m_b$    & $(4.18_{-0.02}^{+0.03})\ $GeV  \cite{ParticleDataGroup:2024cfk}  \\ 
	$m_c$    & $(1.27\pm 0.02)\ $GeV  \cite{ParticleDataGroup:2024cfk}   \\
	$m_{0}^{2}$    & $(0.8\pm0.2)\ $$\mathrm{GeV}^2$  \cite{Belyaev:1982sa,Belyaev:1982cd,Ioffe:2005ym}    
	\\ 
	$\langle \bar{q}q \rangle$    & $-(0.24\pm0.01)^3 $  $\mathrm{GeV}^3$  \cite{Belyaev:1982sa,Belyaev:1982cd}   
	\\ 
	$\langle \bar{s}s \rangle$    & $(0.8\pm0.1) \langle \bar{q}q \rangle $ $\mathrm{GeV}^3$  \cite{Belyaev:1982sa,Belyaev:1982cd}     
	\\ 
	$\langle 0| \frac{1}{\pi} \alpha_s G^2 |0 \rangle $    & $(0.012\pm0.004)\ $ $\mathrm{GeV}^4$  \cite{Belyaev:1982sa,Belyaev:1982cd,Ioffe:2005ym} 
	\\ \hline \hline
\end{tabular}
\end{table}	
\subsection{Working regions of auxiliary parameters}
As a result of the double Borel transformation and continuum subtraction, two auxiliary parameters, namely the Borel parameter $M^2$ and the continuum threshold $s_0$, enter the light cone QCD sum rules. An essential step in the light cone sum rule analysis is to determine the appropriate working regions of these parameters and subsequently fix their values. In the present study, we follow the prescription adopted in Ref.~\cite{Barsbay:2022gtu}. Given the careful analysis of the auxiliary parameters performed in that work, which led to stable predictions for the heavy-light hybrid configurations, and the similarity of the employed QCD sum rule framework to the present analysis, particularly the use of a two point correlation function for determining the relevant hadronic parameters, we adopt the corresponding working regions for the $\mathrm{HM_V}$ and $\mathrm{HM_{AV}}$ heavy-light hybrid mesons. The resulting intervals for the different hybrid configurations are given below. For the ground state charm-nonstrange $\mathrm{HM_V}$ and $\mathrm{HM_{AV}}$ hybrid mesons with $J^{P(C)}=1^{-(-)}$, $1^{-(+)}$, $1^{+(+)}$, and $1^{+(-)}$, the working regions of $M^2$ and $s_0$ are chosen as follows:
\begin{equation}\label{interval1}
	5\ \mathrm{GeV^2} \leq M^2 \leq 7\ \mathrm{GeV^2},\ \     20\, \rm GeV^2 \leq s_0 \leq 24\ \mathrm{GeV^2}  \,. 
\end{equation}
For the ground state charm-strange $\mathrm{HM_V}$ and $\mathrm{HM_{AV}}$ hybrid mesons with $J^{P(C)}=1^{-(-)}$, $1^{-(+)}$, $1^{+(+)}$, and $1^{+(-)}$, the working regions of the Borel parameter $M^2$ and continuum threshold $s_0$ are chosen as,
\begin{equation}\label{interval2}
	5\ \mathrm{GeV^2} \leq M^2 \leq 7\ \mathrm{GeV^2},\ \     22\, \rm GeV^2 \leq s_0 \leq 26\ \mathrm{GeV^2}  \,. 
\end{equation}
Similarly, for the ground state bottom-nonstrange $\mathrm{HM_V}$ and $\mathrm{HM_{AV}}$ hybrid mesons with $J^{P(C)}=1^{-(-)}$, $1^{-(+)}$, $1^{+(+)}$, and $1^{+(-)}$, the corresponding working regions are taken as,
\begin{equation}\label{interval3}
	10\ \mathrm{GeV^2} \leq M^2 \leq 14\ \mathrm{GeV^2},\ \     90\, \rm GeV^2 \leq s_0 \leq 100\ \mathrm{GeV^2}  \,. 
\end{equation}
The above intervals are chosen according to the standard criteria of QCD sum rule analyses. Similar criteria can be applied within the light cone QCD sum rule framework to assess the reliability and appropriateness of the adopted working regions. In the following, we perform a detailed analysis to verify that the selected working regions satisfy the essential requirements of the light cone QCD sum rule method.

First, the continuum threshold $s_0$, which enters as the upper limit of the spectral density integral, is related to the energy of the first excited state and should be chosen such that the ground state contribution is isolated as much as possible from the contributions of higher resonances and the hadronic continuum. This consideration is taken into account in selecting the appropriate intervals of $s_0$ for the different $\mathrm{HM_V}$ and $\mathrm{HM_{AV}}$ hybrid mesons.

Second, the working region of the Borel parameter $M^2$ is determined by imposing two complementary requirements simultaneously: a sufficiently large pole contribution (PC) and a convergent operator product expansion (OPE). The pole dominance condition requires the contribution of the ground state to be sufficiently larger than those of the higher resonances and hadronic continuum. We impose this condition through,
\begin{equation}
	\label{PC}
	\mathrm{PC}=\frac{\Pi^{\rm QCD}(M^{2},\ s_{0})}{\Pi^{\rm QCD}(M^{2},\ \infty )}\,,
\end{equation}
which determines the upper limit of the Borel parameter $M^2$. On the other hand, OPE convergence requires the contributions of higher dimensional condensates as well as higher twists to remain sufficiently small compared with the total QCD contribution. To quantify this requirement, we impose the condition that the contribution of the highest dimensional non-perturbative term included in our OPE, namely the dimension-7 contribution, does not exceed approximately $1\%$ of the total perturbative and non-perturbative photon emission contributions:
\begin{equation}\label{OPEConv}
	\rm OPE\ \ \rm Conv. = \frac{\Pi^{\rm QCD-dim7}(M^2,s_0)}{\Pi^{QCD}(M^2,s_0)} \leq 0.01 \,, 
\end{equation}
where $\Pi^{\rm QCD-dim7}(M^2,s_0)$ represents the contribution of the dimension-7 terms. This criterion determines the lower limit of the Borel parameter $M^2$. The light cone sum rules also exhibit convergence with respect to the contributions of the photon DAs at increasing twist orders.

Third, the physical observables extracted from the light cone sum rules should exhibit sufficient stability and only a weak dependence on the auxiliary parameters within their respective working regions. This stability is essential for ensuring that the obtained predictions are not unduly sensitive to the particular choice of $M^2$ and $s_0$. We examine this requirement in detail below. It is important to emphasize that the working regions of the auxiliary parameters are selected only when all of the above criteria are satisfied simultaneously, thereby ensuring a reliable balance between pole dominance, OPE convergence, and the stability of the physical observables.

\begin{figure}[h!]
	\centering
	\includegraphics[width=0.8\textwidth]{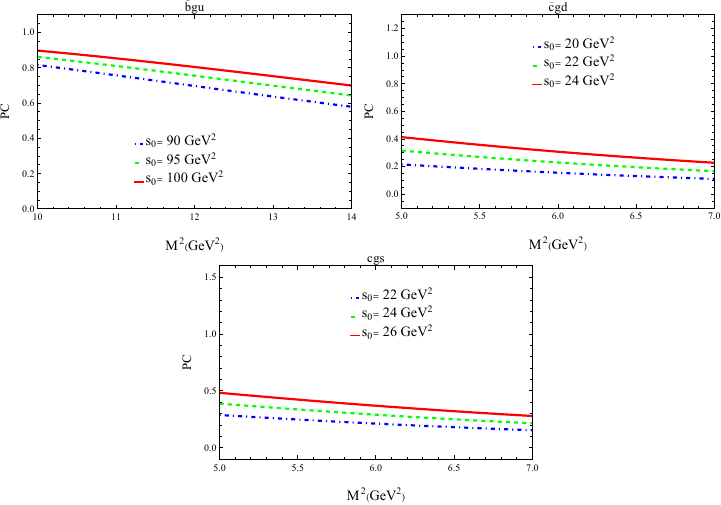}
	\caption{Pole contribution of the $1^{-(-)}$ vector hybrid mesons as a function of $M^2$ and $s_0$.}
	\label{PC1nn}
\end{figure}

\begin{figure}[h!]
	\centering
	\includegraphics[width=0.8\textwidth]{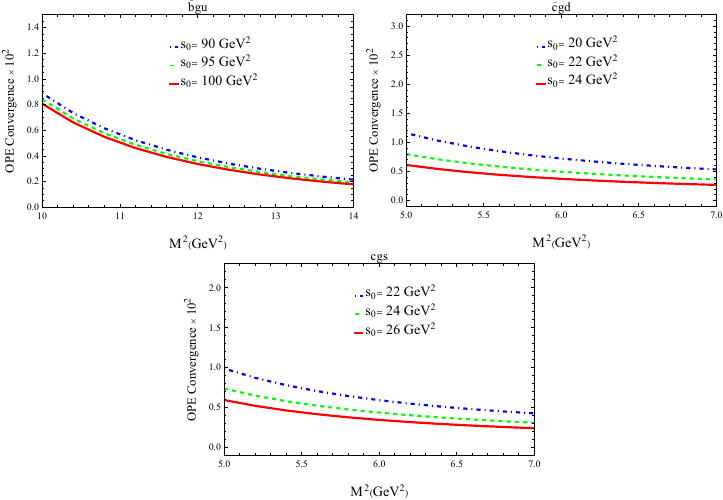}
	\caption{OPE convergence of the $1^{-(-)}$ vector hybrid mesons as a function of $M^2$ and $s_0$.
	}
	\label{OPE1nn}
\end{figure}

\begin{figure}[h!]
	\centering
	\includegraphics[width=0.8\textwidth]{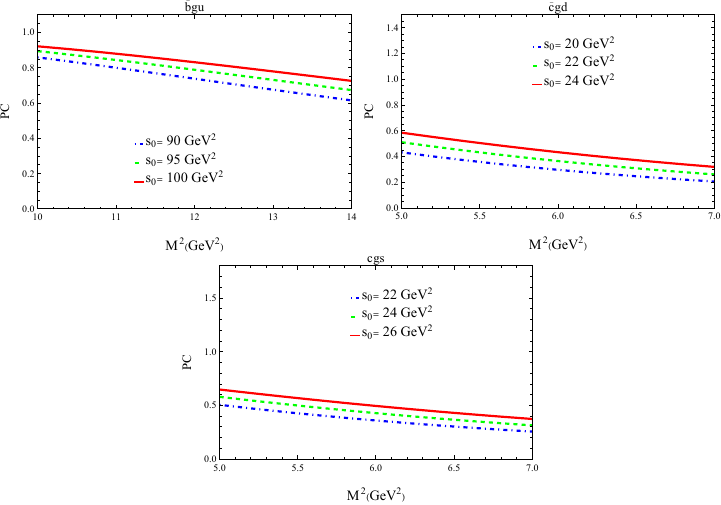}
	\caption{Pole contribution of the $1^{+(-)}$ vector hybrid mesons as a function of $M^2$ and $s_0$.
	}
	\label{PC1pn}
\end{figure}

\begin{figure}[h!]
	\centering
	\includegraphics[width=0.8\textwidth]{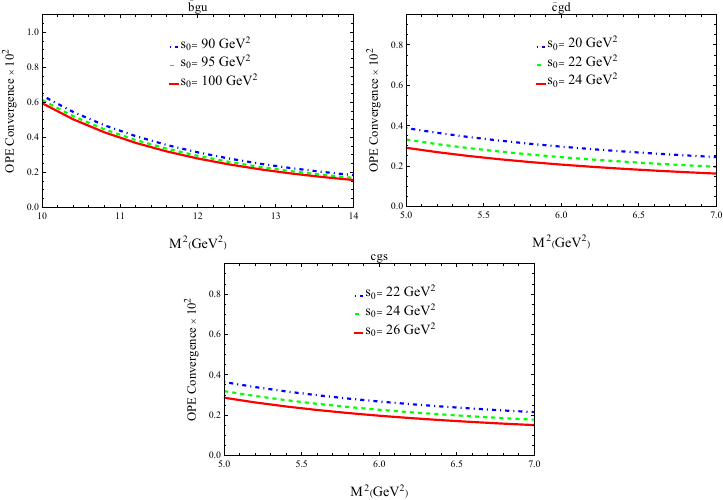}
	\caption{OPE convergence of the $1^{+(-)}$ vector hybrid mesons as a function of $M^2$ and $s_0$.
	}
	\label{OPE1pn}
\end{figure}

In our analysis, we consider the charged heavy-light $\mathrm{HM_V}$ and $\mathrm{HM_{AV}}$ hybrid mesons with the configurations $\bar b g u$, $b g \bar u$, $\bar c g d$, $c g \bar d$, $\bar c g s$, and $c g \bar s$. To assess the reliability of the working regions specified in Eqs.~(\ref{interval1})--(\ref{interval3}), we perform a detailed analysis of the PC and OPE convergence for each hybrid configuration and interpolating current within the corresponding working regions of $M^2$ and $s_0$. The results are summarized in Tables~\ref{PCOPE_results_table1}--\ref{PCOPE_results_table4}. In accordance with the criteria defined in Eqs.~(\ref{PC}) and (\ref{OPEConv}), the continuum threshold $s_0$ is fixed at its central value for each vector and axial-vector hybrid meson, while the Borel parameter $M^2$ is varied throughout its entire working interval. As representative examples, the PC and OPE convergence analyses for the $\bar b g u$, $\bar c g d$, and $\bar c g s$ configurations with quantum numbers $1^{-(-)}$ are presented in Figures~\ref{PC1nn} and~\ref{OPE1nn}, respectively. The corresponding results for the $1^{+(-)}$ states are also shown in Figures~\ref{PC1pn} and~\ref{OPE1pn}. Since the configurations $b g \bar u$, $c g \bar d$, and $c g \bar s$ yield identical results to those of their corresponding charge conjugate configurations, their PC and OPE convergence plots are not presented separately.

As can be seen from the tables and figures, the OPE convergence remains satisfactory for all the considered hybrid states throughout the selected working regions, consistently satisfying the criterion given in Eq.~(\ref{OPEConv}). At the same time, the pole contribution remains sufficiently large within the adopted Borel windows. In particular, at the central values of the $M^2$ and $s_0$ parameters used in the numerical evaluation of the physical observables, the PC is well above $50\%$ for some hybrid states, around or above $40\%$ for others, and around or above $30\%$ for the remaining configurations. These values indicate that the ground state contribution remains sufficiently significant and that the selected working regions provide an acceptable balance between pole dominance and OPE convergence.

A feature that can be observed is that the PC and OPE convergence vary among the different interpolating currents for all considered hybrid configurations. Moreover, the corresponding states with quantum numbers $1^{-(-)}$ and $1^{+(+)}$ exhibit closely related PC and OPE convergence, whereas the states with quantum numbers $1^{-(+)}$ and $1^{+(-)}$ show similarly close behavior.

\begin{figure}[h!]
	\centering
	\includegraphics[width=0.8\textwidth]{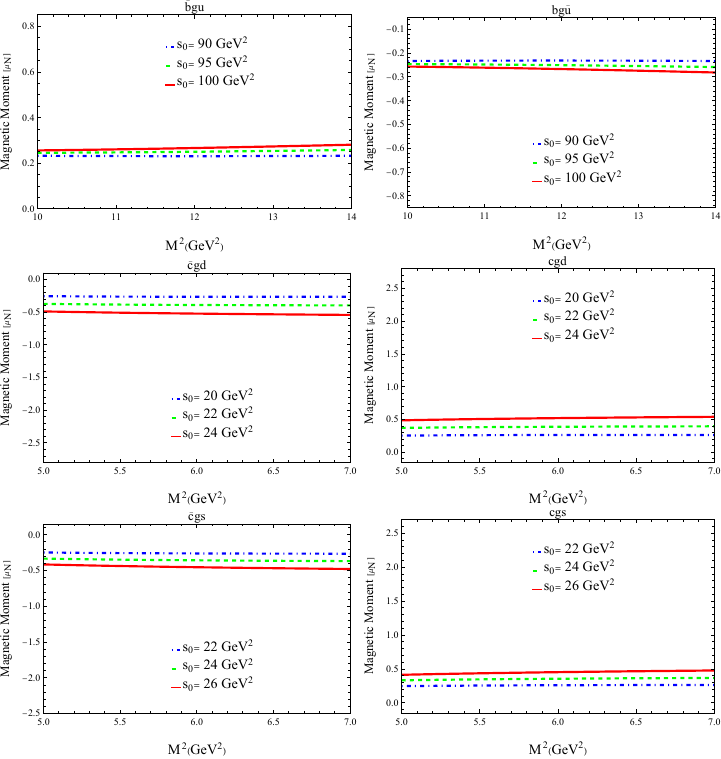}
	\caption{Magnetic moments of the $1^{-(-)}$ vector hybrid mesons as functions of $M^2$ and $s_0$.}
	\label{MM1nn}
\end{figure}

\begin{figure}[h!]
	\centering
	\includegraphics[width=0.8\textwidth]{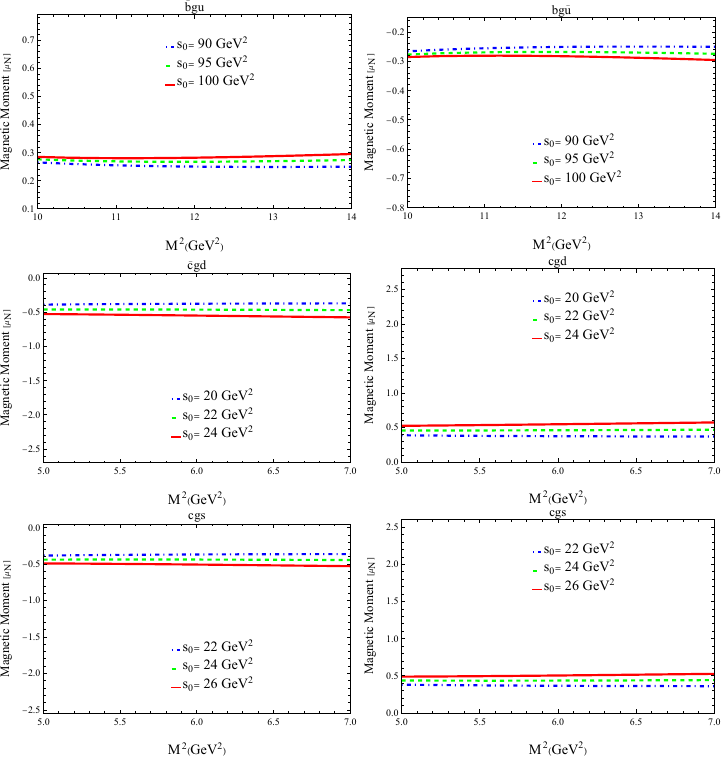}
	\caption{Magnetic moments of the $1^{+(-)}$ vector hybrid mesons as functions of $M^2$ and $s_0$.}
	\label{MM1pn}
\end{figure}

As a final consistency check of the selected working regions, we investigate the stability of the physical observables with respect to variations of the auxiliary parameters. In particular, the dependence of the MDMs on $M^2$ and $s_0$ is illustrated in Figures~\ref{MM1nn} and~\ref{MM1pn} for the $\mathrm{HM_V}$ and $\mathrm{HM_{AV}}$ hybrid mesons with quantum numbers $1^{-(-)}$ and $1^{+(-)}$, respectively. The obtained results exhibit good stability, smooth behavior, and only mild sensitivity to variations of the auxiliary parameters within the selected working regions. Similar behavior is also observed for the remaining $\mathrm{HM_V}$ and $\mathrm{HM_{AV}}$ hybrid configurations and for the other physical observable considered in this work, namely the EQM. 

Overall, the simultaneous fulfillment of the PC and OPE convergence criteria, together with the observed stability of the extracted physical observables, supports the reliability and consistency of the adopted working regions of the auxiliary parameters given in Eqs.~(\ref{interval1})--(\ref{interval3}) for all the $\mathrm{HM_V}$ and $\mathrm{HM_{AV}}$ heavy-light hybrid mesons considered in this study.
\begin{table}[tbp]
	\centering
	\caption{PC and OPE convergence of the heavy-light vector hybrid mesons with $J^{P(C)}=1^{-(-)}$.}
	\label{PCOPE_results_table1}
	\begin{tabular}{|c|c|c|c|c|c|c|}\hline\hline
		$\rm Hybrid\, state$ & $\bar b g u$ & $b g \bar u$  & $\bar c g d$ & $c g \bar d$ & $\bar c g s$ & $c g \bar s$ \\  \hline \hline
		$\rm PC \times 10^2$ & $[86,64]$ &  $[86,64]$ & $[41,23]$ & $[41,23]$ & $[48,28]$ & $[48,20]$\\
		$\rm OPE\, conv. \times 10^2$ & $[0.84,0.19]$ & $[0.84,0.19]$ & $[0.80,0.36]$ & $[0.80,0.36]$ & $[0.74,0.31]$ & $[0.74,0.31]$ \\
		\hline\hline
	\end{tabular}
\end{table}
\begin{table}[tbp]
	\centering
	\caption{PC and OPE convergence of the heavy-light vector hybrid mesons with $J^{P(C)}=1^{-(+)}$.}
	\label{PCOPE_results_table2}
	\begin{tabular}{|c|c|c|c|c|c|c|}\hline\hline
		$\rm Hybrid\, state$ & $\bar b g u$ & $b g \bar u$  & $\bar c g d$ & $c g \bar d$ & $\bar c g s$ & $c g \bar s$ \\  \hline \hline
		$\rm PC \times 10^2$ & $[91,70]$ & $[91,70]$ & $[54,28]$ & $[54,28]$ & $[61,34]$ & $[61,34]$\\
		$\rm OPE\, conv. \times 10^2$ & $[0.49,0.14]$ & $[0.49,0.14]$ & $[0.29,0.17]$ & $[0.29,0.17]$ & $[0.25,0.14]$ & $[0.25,0.14]$ \\
		\hline\hline
	\end{tabular}
\end{table}
\begin{table}[tbp]
	\centering
	\caption{PC and OPE convergence of the heavy-light axial-vector hybrid mesons with $J^{P(C)}=1^{+(+)}$.}
	\label{PCOPE_results_table3}
	\begin{tabular}{|c|c|c|c|c|c|c|}\hline\hline
		$\rm Hybrid\, state$ & $\bar b g u$ & $b g \bar u$  & $\bar c g d$ & $c g \bar d$ & $\bar c g s$ & $c g \bar s$ \\  \hline \hline
		$\rm PC \times 10^2$ & $[88,67]$ &  $[88,67]$ & $[45,25]$ & $[45,25]$ & $[54,31]$ & $[54,31]$\\
		$\rm OPE\, conv. \times 10^2$ & $[0.66,0.16]$ & $[0.66,0.16]$ & $[0.64,0.31]$ & $[0.64,0.31]$ & $[0.53,0.24]$ & $[0.53,0.24]$ \\
		\hline\hline
	\end{tabular}
\end{table}
\begin{table}[tbp]
	\centering
	\caption{PC and OPE convergence of the heavy-light axial-vector hybrid mesons with $J^{P(C)}=1^{+(-)}$.}
	\label{PCOPE_results_table4}
	\begin{tabular}{|c|c|c|c|c|c|c|}\hline\hline
		$\rm Hybrid\, state$ & $\bar b g u$ & $b g \bar u$  & $\bar c g d$ & $c g \bar d$ & $\bar c g s$ & $c g \bar s$ \\  \hline \hline
		$\rm PC \times 10^2$ & $[89,67]$ &  $[89,67]$ & $[51,26]$ & $[51,26]$ & $[58,31]$ & $[58,31]$\\
		$\rm OPE\, conv. \times 10^2$ & $[0.61,0.17]$ & $[0.61,0.17]$ & $[0.33,0.20]$ & $[0.33,0.20]$ & $[0.32,0.18]$ & $[0.32,0.18]$ \\
		\hline\hline
	\end{tabular}
\end{table}
\subsection{Electromagnetic moments}
In this section, using the input parameters and the determined working intervals of $M^2$ and $s_0$, we present the numerical results for the MDMs and EQMs of the $\mathrm{HM_V}$ and $\mathrm{HM_{AV}}$ heavy-light hybrid mesons with different quantum numbers. The results are summarized in Tables~\ref{MM_results_table1}--\ref{MM_results_table4}. The quoted uncertainties arise from the uncertainties of the input parameters, as well as from those associated with the determination of the working regions of the auxiliary parameters. The MDMs and EQMs are given in units of the nuclear magneton and $\rm fm^2$, respectively.

An inspection of Tables~\ref{MM_results_table1}--\ref{MM_results_table4} shows that hybrid states with the same quark and gluon content but different quantum numbers can possess different MDMs and EQMs. This behavior demonstrates the sensitivity of the EM properties not only to the flavor and charge content of the hybrid system, but also to its internal quark-gluon configuration and quantum number assignment. Such a dependence is physically expected, since the MDM probes the spin and magnetization distributions of the constituent degrees of freedom, whereas the EQM is sensitive to the deviation from spherical symmetry in the charge distribution. In a hybrid meson, both quantities are affected by the coupling and spatial organization of the quark and gluonic degrees of freedom. Thus, EM moments provide information complementary to the mass spectrum and other spectroscopic observables and may be sensitive to differences among hybrid configurations containing the same constituent flavors.

A systematic pattern can be observed when the EQMs obtained from the different interpolating currents are compared. The EQMs obtained from $\mathcal{J}_\mu^1$, corresponding to $J^{P(C)}=1^{-(-)}$, and $\mathcal{J}_\mu^3$, corresponding to $J^{P(C)}=1^{+(+)}$, are generally close to each other or approximately compatible within the quoted uncertainties. A similar correspondence is observed between the results associated with $\mathcal{J}_\mu^2$, corresponding to $J^{P(C)}=1^{+(-)}$, and $\mathcal{J}_\mu^4$, corresponding to $J^{P(C)}=1^{-(+)}$. Moreover, the magnitudes of the EQMs associated with the ($\mathcal{J}_\mu^1$,$\mathcal{J}_\mu^3$) group are generally larger than those obtained from the ($\mathcal{J}_\mu^2$,$\mathcal{J}_\mu^4$) group, although the difference is moderate. In contrast to the EQMs, the MDMs do not necessarily exhibit the same systematic grouping among the different hybrid configurations. Nevertheless, the MDMs generally vary when different interpolating currents are employed for the corresponding hybrid states, reflecting their sensitivity to the quantum number and quark-gluon structure encoded in the interpolating currents. Therefore, although the EQMs and MDMs do not necessarily display identical patterns across the different hybrid configurations, their dependence on the interpolating currents demonstrates that the EM response of the hybrid mesons retains information about their underlying quark-gluon configurations. At the same time, although the values of the EM moments vary among the different interpolating currents, these differences are generally moderate.

It is important to emphasize that this observation should not be interpreted as a dependence of a physical observable on the choice of interpolating current itself. Rather, the different currents are employed to interpolate hybrid states with different quantum numbers and internal quark-gluon configurations. The differences among the calculated EM moments therefore reflect the different structures represented by these currents.

Another feature can be observed when comparing the bottom- and charm-containing hybrid states. For the $\bar b g u$ and $b g\bar u$ configurations, the EQMs remain relatively close to one another when the four different interpolating currents are considered. This behavior is less pronounced for the corresponding charm-containing configurations, for which the EQMs exhibit a more noticeable variation among the different currents. This feature is particularly clear in the EQM results, whereas the MDMs also show a similar tendency, although less distinctly. This behavior may be qualitatively related to the heavy quark limit, in which the increasing heavy quark mass can make the heavy quark behave more like a static color source, potentially enhancing the relative role of the light quark and gluonic degrees of freedom in the EM response.

The comparison between the charm and bottom sectors also shows a difference in the magnitudes of the EM moments. For corresponding configurations, the bottom-containing hybrid states generally exhibit smaller MDMs than their charm-containing counterparts. A similar tendency is observed for the EQMs, with the bottom-containing states generally yielding smaller values than the corresponding charm-containing states. These differences may be related, at least qualitatively, to the different heavy quark mass scales in the two sectors. In particular, the larger bottom quark mass can affect the relative contributions of the heavy quark and light-quark/gluonic degrees of freedom to the EM moments. However, the observed differences should not be attributed to a single effect, as the calculated moments also depend on the overall internal structure of the corresponding hybrid states.

The signs of the calculated EM moments provide an additional aspect for comparison among the considered hybrid states. The sign of the MDM reflects the resulting magnetic response of the system, which receives contributions from the charged constituents and their interactions within the hybrid states. The EQM is related to the deviation of the charge distribution from spherical symmetry and can therefore provide information on the non-spherical structure of the hybrid state. In this context, positive and negative EQMs may correspond to different patterns of the charge distribution, which are conventionally associated with prolate-like and oblate-like deformations, respectively. The occurrence of both positive and negative values among the considered configurations therefore indicates differences in the calculated EM properties of the hybrid states. Thus, the MDMs and EQMs offer complementary characteristics of the hybrid states, with the former describing their magnetic response and the latter probing their non-spherical EM structure. The variations in the signs and magnitudes of these moments can consequently be considered when comparing the different heavy-light hybrid configurations.
\begin{table}[tbp]
	\centering
	\caption{Magnetic dipole and quadrupole moments of the vector hybrid mesons with $J^{P(C)}=1^{-(-)}$.}
	\label{MM_results_table1}
	\begin{tabular}{|c|c|c|c|c|c|c|}\hline\hline
		$\rm Hybrid\, State$ & $\bar b g u$ & $b g \bar u$  & $\bar c g d$ & $c g \bar d$ & $\bar c g s$ & $c g \bar s$ \\  \hline \hline
		$\mu_i\, (\mu_N)$ & $0.25_{-0.02}^{+0.03}$ &  $-0.25_{-0.03}^{+0.02}$ & $-0.39_{-0.15}^{+0.13}$ & $0.39_{-0.13}^{+0.15}$ & $-0.35_{-0.12}^{+0.11}$ & $0.35_{-0.11}^{+0.12}$\\
		$\mathcal{D}_i \times 10^{3}\, (fm^2)$ & $-0.35_{-0.09}^{+0.08}$ & $0.35_{-0.08}^{+0.09}$ & $1.17_{-0.24}^{+0.25}$ & $-1.17_{-0.25}^{+0.24}$ & $0.88_{-0.16}^{+0.17}$ & $-0.88_{-0.17}^{+0.16}$ \\
		\hline\hline
	\end{tabular}
\end{table}
\begin{table}[tbp]
	\centering
	\caption{Magnetic dipole and quadrupole moments of the vector hybrid mesons with $J^{P(C)}=1^{-(+)}$.}
	\label{MM_results_table2}
	\begin{tabular}{|c|c|c|c|c|c|c|}\hline\hline
		$\rm Hybrid\, State$ & $\bar b g u$ & $b g \bar u$  & $\bar c g d$ & $c g \bar d$ & $\bar c g s$ & $c g \bar s$ \\  \hline \hline
		$\mu_i\, (\mu_N)$ & $0.32_{-0.03}^{+0.02}$ &  $-0.32_{-0.02}^{+0.03}$ & $-0.52_{-0.11}^{+0.10}$ & $0.52_{-0.10}^{+0.11}$ & $-0.49_{-0.09}^{+0.08}$ & $0.49_{-0.08}^{+0.09}$\\
		$\mathcal{D}_i \times 10^{3}\, (fm^2)$ & $-0.34_{-0.11}^{+0.09}$ & $0.34_{-0.09}^{+0.11}$ & $0.83_{-0.15}^{+0.15}$ & $-0.83_{-0.15}^{+0.15}$ & $0.66_{-0.12}^{+0.13}$ & $-0.66_{-0.13}^{+0.12}$ \\
		\hline\hline
	\end{tabular}
\end{table}
\begin{table}[tbp]
	\centering
	\caption{Magnetic dipole and quadrupole moments of the axial-vector hybrid mesons with $J^{P(C)}=1^{+(+)}$.}
	\label{MM_results_table3}
	\begin{tabular}{|c|c|c|c|c|c|c|}\hline\hline
		$\rm Hybrid\, State$ & $\bar b g u$ & $b g \bar u$  & $\bar c g d$ & $c g \bar d$ & $\bar c g s$ & $c g \bar s$ \\  \hline \hline
		$\mu_i\, (\mu_N)$ & $0.29_{-0.02}^{+0.03}$ &  $-0.29_{-0.03}^{+0.02}$ & $-0.44_{-0.16}^{+0.13}$ & $0.44_{-0.13}^{+0.16}$ & $-0.44_{-0.12}^{+0.10}$ & $0.44_{-0.10}^{+0.12}$\\
		$\mathcal{D}_i \times 10^{3}\, (fm^2)$ & $-0.33_{-0.09}^{+0.07}$ & $0.33_{-0.07}^{+0.09}$ & $1.04_{-0.18}^{+0.18}$ & $-1.04_{-0.18}^{+0.18}$ & $0.81_{-0.14}^{+0.14}$ & $-0.81_{-0.14}^{+0.14}$ \\
		\hline\hline
	\end{tabular}
\end{table}
\begin{table}[tbp]
	\centering
	\caption{Magnetic dipole and quadrupole moments of the axial-vector hybrid mesons with $J^{P(C)}=1^{+(-)}$.}
	\label{MM_results_table4}
	\begin{tabular}{|c|c|c|c|c|c|c|}\hline\hline
		$\rm Hybrid\, State$ & $\bar b g u$ & $b g \bar u$  & $\bar c g d$ & $c g \bar d$ & $\bar c g s$ & $c g \bar s$ \\  \hline \hline
		$\mu_i\, (\mu_N)$ & $0.27_{-0.02}^{+0.02}$ &  $-0.27_{-0.02}^{+0.02}$ & $-0.46_{-0.11}^{+0.09}$ & $0.46_{-0.09}^{+0.11}$ & $-0.44_{-0.08}^{+0.07}$ & $0.44_{-0.07}^{+0.08}$\\
		$\mathcal{D}_i \times 10^{3}\, (fm^2)$ & $-0.35_{-0.11}^{+0.09}$ & $0.35_{-0.09}^{+0.11}$ & $0.83_{-0.14}^{+0.15}$ & $-0.83_{-0.15}^{+0.14}$ & $0.72_{-0.13}^{+0.14}$ & $-0.72_{-0.14}^{+0.13}$ \\
		\hline\hline
	\end{tabular}
\end{table}
\section{Conclusions}\label{sec:conclusions}
In this work, we have investigated the electromagnetic properties of vector and axial-vector heavy-light hybrid mesons containing an explicit gluonic degree of freedom. In particular, their magnetic dipole and electric quadrupole moments have been calculated within the light cone QCD sum rules framework. Four different interpolating currents corresponding to the $J^{P(C)}=1^{-(-)}$, $1^{+(-)}$, $1^{+(+)}$, and $1^{-(+)}$ hybrid configurations have been considered, allowing us to investigate the electromagnetic properties of the corresponding hybrid states.

The light cone sum rules of the magnetic dipole and electric quadrupole moments have been obtained by matching the hadronic and QCD representations of the correlation function, followed by the appropriate double Borel transformation and continuum subtraction. In the QCD representation, both perturbative and non-perturbative photon emission contributions have been taken into account, with the latter including the relevant photon distribution amplitudes of different twists. The numerical analysis provides predictions for the electromagnetic moments of the charged $\bar b g u$, $b g\bar u$, $\bar c g d$, $c g\bar d$, $\bar c g s$, and $c g\bar s$ hybrid configurations.

Our numerical results show that the calculated electromagnetic moments vary among the different hybrid configurations. For states with the same quark and gluon content but different quantum numbers, different values of the magnetic dipole and electric quadrupole moments are generally obtained. A systematic pattern is more clearly observed for the quadrupole moments. The results associated with the $\mathcal J_\mu^1$ and $\mathcal J_\mu^3$ currents are generally close to each other or approximately compatible within the quoted uncertainties, while a similar correspondence is observed between the $\mathcal J_\mu^2$ and $\mathcal J_\mu^4$ currents. Moreover, the magnitudes of the quadrupole moments associated with the ($\mathcal J_\mu^1$,$\mathcal J_\mu^3$) group are generally larger than those obtained from the ($\mathcal J_\mu^2$,$\mathcal J_\mu^4$) group, although the difference is moderate. The magnetic dipole moments, in contrast, do not necessarily exhibit the same systematic grouping, although they also vary among the different interpolating currents and corresponding hybrid configurations. These results indicate that the calculated EM moments can differ according to the quantum numbers and quark-gluon structures represented by the interpolating currents.

A further feature is observed when comparing the bottom- and charm-containing hybrid states. For the $\bar b g u$ and $b g\bar u$ configurations, the quadrupole moments remain relatively close when the four different interpolating currents are considered, whereas this behavior is less pronounced for the corresponding charm-containing configurations, for which the quadrupole moments show a more noticeable variation among the different currents. This feature is particularly clear in the quadrupole moment results, while the magnetic dipole moments show a similar tendency, although less distinctly. Such differences may be related, at least qualitatively, to the different heavy quark mass scales in the charm and bottom sectors. We also find that the magnitudes of the magnetic dipole and quadrupole moments of the bottom-containing hybrid states are generally smaller than those of the corresponding charm-containing states. These observations indicate a dependence of the calculated EM moments on the heavy flavor content, although the detailed origin of the differences cannot be attributed to a single effect.

The signs of the calculated moments provide an additional aspect for comparison among the considered hybrid states. The sign of the magnetic dipole moment reflects the resulting magnetic response of the system, while the electric quadrupole moment is related to the non-spherical part of the charge distribution. The occurrence of both positive and negative values among the considered configurations therefore points to differences in the corresponding EM properties. In this sense, the magnetic dipole and electric quadrupole moments provide complementary characteristics of the hybrid states, with the former describing their magnetic response and the latter being associated with their non-spherical EM structure. The variations in the signs and magnitudes of these moments can consequently be considered when comparing the different heavy-light hybrid configurations.

Overall, the results obtained in this work provide predictions for the magnetic dipole and electric quadrupole moments of heavy-light hybrid mesons and show variations associated with their quantum numbers and heavy flavor content. In particular, the quadrupole moments exhibit identifiable patterns among the different interpolating currents, while the magnetic dipole moments show a less systematic behavior. These results provide complementary information on the EM properties of heavy-light hybrid mesons and may be useful for future theoretical studies of their structure and, where experimentally accessible, for investigations of their EM properties.

\appendix
\section{Distribution Amplitudes of the on-shell photon} \label{AppenPDAs}
In this appendix, we collect the matrix elements $\langle \gamma(q)| \bar{q}(x) \Gamma_i G_{\mu\nu}q(0) | 0\rangle$ and $\langle \gamma(q)| \bar{q}(x) \Gamma_i q(0) | 0\rangle$ that enter the non-perturbative part of the QCD representation. These matrix elements are expressed in terms of the photon DAs introduced below~\cite{Ball:2002ps}:
\begin{eqnarray}
	\label{esbs14}
	&&\langle \gamma(q) \vert  \bar q(x) \gamma_\mu q(0) \vert 0 \rangle
	= e_q f_{3 \gamma} \left(\varepsilon_\mu - q_\mu \frac{\varepsilon
		x}{q x} \right) \int_0^1 du e^{i \bar u q x} \psi^v(u)
	\nonumber \\
	&&\langle \gamma(q) \vert \bar q(x) \gamma_\mu \gamma_5 q(0) \vert 0
	\rangle  = - \frac{1}{4} e_q f_{3 \gamma} \epsilon_{\mu \nu \alpha
		\beta } \varepsilon^\nu q^\alpha x^\beta \int_0^1 du e^{i \bar u q
		x} \psi^a(u)
	\nonumber \\
	&&\langle \gamma(q) \vert  \bar q(x) \sigma_{\mu \nu} q(0) \vert  0
	\rangle  = -i e_q \langle \bar q q \rangle (\varepsilon_\mu q_\nu - \varepsilon_\nu
	q_\mu) \int_0^1 du e^{i \bar u qx} \left(\chi \varphi_\gamma(u) +
	\frac{x^2}{16} \mathbb{A}  (u) \right) \nonumber \\ 
	&&-\frac{i}{2(qx)}  e_q \langle \bar q q \rangle \left[x_\nu \left(\varepsilon_\mu - q_\mu
	\frac{\varepsilon x}{qx}\right) - x_\mu \left(\varepsilon_\nu -
	q_\nu \frac{\varepsilon x}{q x}\right) \right] \int_0^1 du e^{i \bar
		u q x} h_\gamma(u)
	\nonumber \\
	&&\langle \gamma(q) | \bar q(x) g_s G_{\mu \nu} (v x) q(0) \vert 0
	\rangle = -i e_q \langle \bar q q \rangle \left(\varepsilon_\mu q_\nu - \varepsilon_\nu
	q_\mu \right) \int {\cal D}\alpha_i e^{i (\alpha_{\bar q} + v
		\alpha_g) q x} {\cal S}(\alpha_i)
	\nonumber \\
	&&\langle \gamma(q) | \bar q(x) g_s \tilde G_{\mu \nu}(v
	x) i \gamma_5  q(0) \vert 0 \rangle = -i e_q \langle \bar q q \rangle \left(\varepsilon_\mu q_\nu -
	\varepsilon_\nu q_\mu \right) \int {\cal D}\alpha_i e^{i
		(\alpha_{\bar q} + v \alpha_g) q x} \tilde {\cal S}(\alpha_i)
	\nonumber \\
	&&\langle \gamma(q) \vert \bar q(x) g_s \tilde G_{\mu \nu}(v x)
	\gamma_\alpha \gamma_5 q(0) \vert 0 \rangle = e_q f_{3 \gamma}
	q_\alpha (\varepsilon_\mu q_\nu - \varepsilon_\nu q_\mu) \int {\cal
		D}\alpha_i e^{i (\alpha_{\bar q} + v \alpha_g) q x} {\cal
		A}(\alpha_i)
	\nonumber \\
	&&\langle \gamma(q) \vert \bar q(x) g_s G_{\mu \nu}(v x) i
	\gamma_\alpha q(0) \vert 0 \rangle = e_q f_{3 \gamma} q_\alpha
	(\varepsilon_\mu q_\nu - \varepsilon_\nu q_\mu) \int {\cal
		D}\alpha_i e^{i (\alpha_{\bar q} + v \alpha_g) q x} {\cal
		V}(\alpha_i) \nonumber\\
	&& \langle \gamma(q) \vert \bar q(x)
	\sigma_{\alpha \beta} g_s G_{\mu \nu}(v x) q(0) \vert 0 \rangle  =
	e_q \langle \bar q q \rangle \left\{
	\left[\left(\varepsilon_\mu - q_\mu \frac{\varepsilon x}{q x}\right)\left(g_{\alpha \nu} -
	\frac{1}{qx} (q_\alpha x_\nu + q_\nu x_\alpha)\right) \right. \right. q_\beta
	\nonumber \\
	&& -
	\left(\varepsilon_\mu - q_\mu \frac{\varepsilon x}{q x}\right)\left(g_{\beta \nu} -
	\frac{1}{qx} (q_\beta x_\nu + q_\nu x_\beta)\right) q_\alpha
	-
	\left(\varepsilon_\nu - q_\nu \frac{\varepsilon x}{q x}\right)\left(g_{\alpha \mu} -
	\frac{1}{qx} (q_\alpha x_\mu + q_\mu x_\alpha)\right) q_\beta
	\nonumber \\
	&&+
	\left. \left(\varepsilon_\nu - q_\nu \frac{\varepsilon x}{qx}\right)\left( g_{\beta \mu} -
	\frac{1}{qx} (q_\beta x_\mu + q_\mu x_\beta)\right) q_\alpha \right]
	\int {\cal D}\alpha_i e^{i (\alpha_{\bar q} + v \alpha_g) qx} {\cal T}_1(\alpha_i)
	\nonumber \\
	&&+
	\left[\left(\varepsilon_\alpha - q_\alpha \frac{\varepsilon x}{qx}\right)
	\left(g_{\mu \beta} - \frac{1}{qx}(q_\mu x_\beta + q_\beta x_\mu)\right) \right. q_\nu -
	\left(\varepsilon_\alpha - q_\alpha \frac{\varepsilon x}{qx}\right)
	\left(g_{\nu \beta} - \frac{1}{qx}(q_\nu x_\beta + q_\beta x_\nu)\right)  q_\mu
	\nonumber \\ && -
	\left(\varepsilon_\beta - q_\beta \frac{\varepsilon x}{qx}\right)
	\left(g_{\mu \alpha} - \frac{1}{qx}(q_\mu x_\alpha + q_\alpha x_\mu)\right) q_\nu
	\nonumber \\ &&+
	\left. \left(\varepsilon_\beta - q_\beta \frac{\varepsilon x}{qx}\right)
	\left(g_{\nu \alpha} - \frac{1}{qx}(q_\nu x_\alpha + q_\alpha x_\nu) \right) q_\mu
	\right]      
	\int {\cal D} \alpha_i e^{i (\alpha_{\bar q} + v \alpha_g) qx} {\cal T}_2(\alpha_i)
	\nonumber \\
&&+\frac{1}{qx} (q_\mu x_\nu - q_\nu x_\mu)
(\varepsilon_\alpha q_\beta - \varepsilon_\beta q_\alpha)
\int {\cal D} \alpha_i e^{i (\alpha_{\bar q} + v \alpha_g) qx} {\cal T}_3(\alpha_i)
\nonumber \\ &&+
\left. \frac{1}{qx} (q_\alpha x_\beta - q_\beta x_\alpha)
(\varepsilon_\mu q_\nu - \varepsilon_\nu q_\mu)
\int {\cal D} \alpha_i e^{i (\alpha_{\bar q} + v \alpha_g) qx} {\cal T}_4(\alpha_i)
\right\}\,.
\end{eqnarray}
The integration measure $\mathcal{D} \alpha_i$ appearing above is given by,
\begin{eqnarray}
\label{nolabel05}
\int {\cal D} \alpha_i = \int_0^1 d \alpha_{\bar q} \int_0^1 d \alpha_q \int_0^1 d \alpha_g \delta(1-\alpha_{\bar q}-\alpha_q-\alpha_g)~\,.
\end{eqnarray}
In the above matrix elements, the leading twist-2 contribution is described by the photon DA $\varphi_\gamma(u)$. At twist-3 accuracy, the relevant photon DAs are $\psi^v(u)$, $\psi^a(u)$, ${\cal A}(\alpha_i)$, and ${\cal V}(\alpha_i)$. The twist-4 sector, in turn, involves the functions $h_\gamma(u)$, $\mathbb{A}(u)$, ${\cal S}(\alpha_i)$, ${\cal \tilde{S}}(\alpha_i)$, ${\cal T}_1(\alpha_i)$, ${\cal T}_2(\alpha_i)$, ${\cal T}_3(\alpha_i)$, and ${\cal T}_4(\alpha_i)$.

The photon DAs entering the matrix elements above are specified by the following expressions:
\begin{eqnarray}
\varphi_\gamma(u) &=& 6 u \bar u \left( 1 + \varphi_2(\mu)
C_2^{\frac{3}{2}}(u - \bar u) \right),
\nonumber \\
\psi^v(u) &=& 5 \left(3 (2 u - 1)^2 -1 \right)+\frac{3}{64} \left(15
w^V_\gamma - 5 w^A_\gamma\right)
\left(3 - 30 (2 u - 1)^2 + 35 (2 u -1)^4
\right),
\nonumber \\
\psi^a(u) &=& \left(1- (2 u -1)^2\right)\left(5 (2 u -1)^2 -1\right)
\frac{5}{2}
\left(1 + \frac{9}{16} w^V_\gamma - \frac{3}{16} w^A_\gamma
\right),
\nonumber \\
h_\gamma(u) &=& - 10 \left(1 + 2 \kappa^+\right) C_2^{\frac{1}{2}}(u
- \bar u),
\nonumber \\
\mathbb{A}(u) &=& 40 u^2 \bar u^2 \left(3 \kappa - \kappa^+
+1\right)  +
8 (\zeta_2^+ - 3 \zeta_2) \left[u \bar u (2 + 13 u \bar u) \right. \nonumber \\ &&+ \left.
2 u^3 (10 -15 u + 6 u^2) \ln(u) 
 + 2 \bar u^3 (10 - 15 \bar u + 6 \bar u^2) \ln(\bar u) \right],
\nonumber \\
{\cal A}(\alpha_i) &=& 360 \alpha_q \alpha_{\bar q} \alpha_g^2
\left(1 + w^A_\gamma \frac{1}{2} (7 \alpha_g - 3)\right),
\nonumber \\
{\cal V}(\alpha_i) &=& 540 w^V_\gamma (\alpha_q - \alpha_{\bar q})
\alpha_q \alpha_{\bar q}
\alpha_g^2,
\nonumber \\
{\cal T}_1(\alpha_i) &=& -120 (3 \zeta_2 + \zeta_2^+)(\alpha_{\bar
	q} - \alpha_q)
\alpha_{\bar q} \alpha_q \alpha_g,
\nonumber \\
{\cal T}_2(\alpha_i) &=& 30 \alpha_g^2 (\alpha_{\bar q} - \alpha_q)
\left((\kappa - \kappa^+) + (\zeta_1 - \zeta_1^+)(1 - 2\alpha_g) +
\zeta_2 (3 - 4 \alpha_g)\right),
\nonumber \\
{\cal T}_3(\alpha_i) &=& - 120 (3 \zeta_2 - \zeta_2^+)(\alpha_{\bar
	q} -\alpha_q)
\alpha_{\bar q} \alpha_q \alpha_g,
\nonumber \\
{\cal T}_4(\alpha_i) &=& 30 \alpha_g^2 (\alpha_{\bar q} - \alpha_q)
\left((\kappa + \kappa^+) + (\zeta_1 + \zeta_1^+)(1 - 2\alpha_g) +
\zeta_2 (3 - 4 \alpha_g)\right),\nonumber \\
{\cal S}(\alpha_i) &=& 30\alpha_g^2\{(\kappa +
\kappa^+)(1-\alpha_g)+(\zeta_1 + \zeta_1^+)(1 - \alpha_g)(1 -
2\alpha_g)\nonumber +\zeta_2[3 (\alpha_{\bar q} - \alpha_q)^2-\alpha_g(1 - \alpha_g)]\},\nonumber \\
\tilde {\cal S}(\alpha_i) &=&-30\alpha_g^2\{(\kappa -\kappa^+)(1-\alpha_g)+(\zeta_1 - \zeta_1^+)(1 - \alpha_g)(1 -
2\alpha_g) +\zeta_2 [3 (\alpha_{\bar q} -\alpha_q)^2-\alpha_g(1 - \alpha_g)]\}\,. \nonumber \\ 
\end{eqnarray}
For the numerical analysis, we employ the following values for the parameters entering the matrix elements and photon DAs: $\chi(1\,\rm GeV) =-2.85\pm 0.5 \rm\, GeV^{-2}$~\cite{Rohrwild:2007yt}, $f_{3\gamma}=-0.0039\, \rm GeV^2$, $\varphi_2(1\,\rm GeV) = 0$, 
$w^V_\gamma = 3.8 \pm 1.8$, $w^A_\gamma = -2.1 \pm 1.0$, $\kappa = 0.2$, $\kappa^+ = 0$, $\zeta_1 = 0.4$, $\zeta_2 = 0.3$, $\zeta_1^+ = 0$, and $\zeta_2^+ = 0$.

\section{Expressions of QCD correlation function} \label{AppenCorQCD}
In this appendix, we present, as an illustrative example, the QCD correlation functions $\Pi^{\rm QCD}(\rm M^2,s_0)$ and $\Pi^{\prime \rm QCD}(\rm{M^2},\rm{s_0})$ entering the light cone sum rules for the vector hybrid mesons with $J^{P(C)}=1^{-(-)}$. The first correlation function, $\Pi^{\rm QCD}(\rm M^2,s_0)$, is given by,
\begin{align}\label{CorrFpertQCD}
	\Pi^{\mathrm{QCD}}&(\rm{M^2},\rm{s_0})= 
	\frac{g_s^2}{4608\,m_Q^2\pi^4} \Bigg\{-3 e_q m_Q^5 \Bigg[-12 m_q \Big(I_{ls}[1,1]+I_{ls}[2,2] +m_Q^4I_{ls}[3,1]-m_Q^2\big(2I_{ls}[2,1]+I_{ls}[3,2]\big) \Big) \nonumber\\&
	+6m_q^2\, m_Q \Big(-I_{ls}[2,1]-I_{ls}[3,2] + m_Q^2\big(I_{ls}[3,1]+I_{ls}[4,2]\big) \Big)
	\nonumber\\& + m_Q \Big(3I_{ls}[2,2]+5I_{ls}[3,3] -6m_Q^2\big(I_{ls}[3,2]+I_{ls}[4,3]\big)
	+m_Q^4\big(3I_{ls}[4,2]+I_{ls}[5,3]\big) \Big) \Bigg] \nonumber\\&
	+2\, e_Q \Bigg[-3m_Q^3 \Big(8\pi^2\langle\bar qq\rangle \big(I_{ls}[0,0]+I_{ls}[1,1]\big)
	-m_q\big(2I_{ls}[0,1]+3I_{ls}[1,2]\big)\Big) \nonumber\\&
	+6m_Q^5 \Big(8\pi^2\langle\bar qq\rangle I_{ls}[1,0] -3m_q\big(I_{ls}[1,1]+I_{ls}[2,2]\big)\Big)
	-6m_qm_Q^9I_{ls}[3,1] \nonumber\\&
	+3m_Q^7 \Big( 8\pi^2\langle\bar qq\rangle \big(-I_{ls}[2,0]+I_{ls}[3,1]\big)
	+3m_q\big(2I_{ls}[2,1]+I_{ls}[3,2]\big) \Big) \nonumber\\&
	+3m_Q^6 \Big(3I_{ls}[2,2]+8m_q\pi^2\langle\bar qq\rangle \big(I_{ls}[2,0]-I_{ls}[3,1]\big) +5I_{ls}[3,3]\Big) \nonumber\\&
	-9m_Q^8\big(I_{ls}[3,2]+I_{ls}[4,3]\big) +m_Q^{10}\big(3I_{ls}[4,2]+I_{ls}[5,3]\big) \nonumber\\&
	-m_Q^4 \Big(3I_{ls}[1,2]+7I_{ls}[2,3] +24m_q\pi^2\langle\bar qq\rangle \big(I_{ls}[1,0]+I_{ls}[2,1]+I_s[0]\big) \Big) \nonumber\\&
	+48m_qm_Q^2\pi^2\langle\bar qq\rangle I_s[1] + 24m_q\pi^2\langle\bar qq\rangle \big(2I_{ls}[0,1]-I_s[2]\big) \nonumber\\&
	+4m_0^2\pi^2\langle\bar qq\rangle \Bigg(m_q\Big(2I_{ls}[0,0]+m_Q^4\big(-2I_{ls}[2,0]+I_s[-1]\big)
	+4m_Q^2I_s[0]-2I_s[1]\Big)\nonumber\\&
	+3m_Q\Big(-I_{ls}[0,0]-m_Q^2I_{ls}[1,0]+m_Q^4\big(2I_{ls}[2,0]-I_s[-1]\big)+I_s[1]\Big)\Bigg)\Bigg]
	\Bigg\} \nonumber\\&  
	+ \frac{\langle g_s^2G^2 \rangle}{13824\,m_Q^2M^6\pi^2} e^{-m_Q^2/M^2}
	\Bigg\{-4e_Q\Bigg[m_0^2m_Q^2 \Big(-3m_Q^3M^2
	+m_q\big(m_Q^4+8m_Q^2M^2-8M^4\big)\Big) \pi^2 \langle\bar qq\rangle \nonumber\\&
	+3M^4\Bigg(m_qm_Q\Big(2m_Q^3\pi^2 \langle\bar qq\rangle +10m_QM^2\pi^2 \langle\bar qq\rangle
	+3e^{m_Q^2/M^2}m_Q^2M^2 \big(I_{ls}[1,0]-I_s[0]\big) \nonumber\\&
	-3e^{m_Q^2/M^2}M^2 \big(I_{ls}[0,0]-I_s[1]\big)\Big)
	-M^2\Big(4m_Q^3\pi^2 \langle\bar qq\rangle 
	-3e^{m_Q^2/M^2}m_Q^6 \big(I_{ls}[2,0]-I_{ls}[3,1]\big) \nonumber\\&
	+3e^{m_Q^2/M^2}m_Q^4 \big(I_{ls}[1,0]+I_{ls}[2,1]+I_s[0]\big)
	-6e^{m_Q^2/M^2}m_Q^2I_s[1]-3e^{m_Q^2/M^2} \big(2I_{ls}[0,1]-I_s[2]\big)\Big)\Bigg)\Bigg]\nonumber\\&
	+9e_qM^6\Bigg(m_q^2m_Q^4-4e^{m_Q^2/M^2}m_qm_Q^3I_s[0] + e^{m_Q^2/M^2}\Big(2I_{ls}[0,1]
	+2m_Q^6\big(I_{ls}[2,0]-I_{ls}[3,1]\big) \nonumber\\& -7m_Q^4 I_s[0]
	-2m_Q^2\big(I_{ls}[0,0]-4I_s[1]\big) -I_s[2]\Big)\Bigg)\Bigg\} \nonumber\\& 
	+ \frac{e_q g_s^2}{4608\,m_Q^2\pi^2} \Bigg\{ -6m_Q\langle\bar q q\rangle \mathbb{A}[u_0]
	\Big(-2I_{ls}[0,1] +4m_Q^6I_{ls}[3,1] + m_Q^4\big(-2I_{ls}[2,1]+I_s[0]\big) \nonumber\\& -2m_Q^2 I_s[1]
	+I_s[2] \Big)  -24\chi m_Q^7\langle\bar q q\rangle \big(-I_{ls}[3,2]+m_Q^2I_{ls}[4,2]\big)
	\phi_\gamma[u_0] -24f_{3\gamma}I_{ls}[0,2]\psi^a[u_0] \nonumber\\&
	+24f_{3\gamma}m_Q^8 I_{ls}[4,2]\psi^a[u_0] -8f_{3\gamma}m_Q^6 I_s[0]\psi^a[u_0] +24f_{3\gamma}m_Q^4 I_s[1]\psi^a[u_0] -24 f_{3\gamma}m_Q^2 I_s[2]\psi^a[u_0] \nonumber\\&
	+8f_{3\gamma} I_s[3] \psi^a[u_0] -24 f_{3\gamma}m_Q^6 I_{ls}[3,2] \psi^v[u_0] +24 f_{3\gamma} m_Q^8 I_{ls}[4,2]\psi^v [u_0] - 24 m_Q^5 \langle\bar q q\rangle I_{ls}[2,1] I_1[h_\gamma,u]
	\nonumber\\&
    +24m_Q^7\langle\bar q q\rangle I_{ls}[3,1]I_1[h_\gamma,u] +48f_{3\gamma}m_Q^6I_{ls}[3,2]I_1[\psi^v,u] -48f_{3\gamma}m_Q^8 I_{ls}[4,2] I_1[\psi^v,u] \nonumber\\&
	-48m_Q^5\langle\bar q q\rangle I_{ls}[2,1] I_2[h_\gamma,u]
	+48m_Q^7\langle\bar q q\rangle I_{ls}[3,1] I_2[h_\gamma,u]
	-132 m_Q^3\langle\bar q q\rangle I_{ls}[1,1] I_{3}[\mathcal{S},\alpha_i] \nonumber\\&
	+528 m_Q^5\langle\bar q q\rangle I_{ls}[2,1] I_{3}[\mathcal{S},\alpha_i]-396 m_Q^7\langle\bar q q\rangle I_{ls}[3,1] I_{3}[\mathcal{S},\alpha_i] -66 m_Q^3\langle\bar q q\rangle I_{ls}[1,1] I_{3}[\mathcal{T}_1,\alpha_i]  \nonumber\\&
	+66 m_Q^7\langle\bar q q\rangle I_{ls}[3,1] I_{3}[\mathcal{T}_1,\alpha_i]
	-66 m_Q^3\langle\bar q q\rangle I_{ls}[1,1] I_{3}[\mathcal{T}_2,\alpha_i] +66 m_Q^7\langle\bar q q\rangle I_{ls}[3,1] I_{3}[\mathcal{T}_2,\alpha_i] \nonumber\\&
	+66 m_Q^3\langle\bar q q\rangle I_{ls}[1,1] I_{3}[\mathcal{T}_3,\alpha_i] -66 m_Q^7\langle\bar q q\rangle I_{ls}[3,1] I_{3}[\mathcal{T}_3,\alpha_i]+66 m_Q^3\langle\bar q q\rangle I_{ls}[1,1] I_{3}[\mathcal{T}_4,\alpha_i] \nonumber\\& -66 m_Q^7\langle\bar q q\rangle I_{ls}[3,1] I_{3}[\mathcal{T}_4,\alpha_i]
	-132 m_Q^3\langle\bar q q\rangle I_{ls}[1,1] I_{3}[\tilde{\mathcal{S}},\alpha_i] +528 m_Q^5\langle\bar q q\rangle I_{ls}[2,1] I_{3}[\tilde{\mathcal{S}},\alpha_i] \nonumber\\&
	-396 m_Q^7\langle\bar q q\rangle I_{ls}[3,1] I_{3}[\tilde{\mathcal{S}},\alpha_i] +99 f_{3\gamma}m_Q^4 I_{ls}[2,2] I_4[\mathcal{A},\alpha_i] 
	-132 f_{3\gamma}m_Q^6 I_{ls}[3,2] I_4[\mathcal{A},\alpha_i] \nonumber\\& + 33 f_{3\gamma} m_Q^8 I_{ls}[4,2] I_4[\mathcal{A},\alpha_i] +66 m_Q^3\langle\bar q q\rangle I_{ls}[1,1] I_4[\mathcal{S},\alpha_i] -132 m_Q^5\langle\bar q q\rangle I_{ls}[2,1] I_4[\mathcal{S},\alpha_i] \nonumber\\&
	+66 m_Q^7\langle\bar q q\rangle I_{ls}[3,1] I_4[\mathcal{S},\alpha_i] -33 m_Q^3\langle\bar q q\rangle I_{ls}[1,1] I_4[\mathcal{T}_1,\alpha_i] 
	+66 m_Q^5\langle\bar q q\rangle I_{ls}[2,1] I_4[\mathcal{T}_1,\alpha_i] \nonumber\\& 
	-33 m_Q^7\langle\bar q q\rangle I_{ls}[3,1] I_4[\mathcal{T}_1,\alpha_i]
	-33 m_Q^3\langle\bar q q\rangle I_{ls}[1,1] I_4[\mathcal{T}_2,\alpha_i] +66 m_Q^5\langle\bar q q\rangle I_{ls}[2,1] I_4[\mathcal{T}_2,\alpha_i] \nonumber\\& 
	-33 m_Q^7\langle\bar q q\rangle I_{ls}[3,1] I_4[\mathcal{T}_2,\alpha_i] +33 m_Q^3\langle\bar q q\rangle I_{ls}[1,1] I_4[\mathcal{T}_3,\alpha_i] -66 m_Q^5\langle\bar q q\rangle I_{ls}[2,1] I_4[\mathcal{T}_3,\alpha_i] \nonumber\\& 
	+33 m_Q^7\langle\bar q q\rangle I_{ls}[3,1] I_4[\mathcal{T}_3,\alpha_i] 
	+33 m_Q^3\langle\bar q q\rangle I_{ls}[1,1] I_4[\mathcal{T}_4,\alpha_i] -66 m_Q^5\langle\bar q q\rangle I_{ls}[2,1] I_4[\mathcal{T}_4,\alpha_i] \nonumber\\&
	+33 m_Q^7\langle\bar q q\rangle I_{ls}[3,1] I_4[\mathcal{T}_4,\alpha_i] +66 m_Q^3\langle\bar q q\rangle I_{ls}[1,1] I_4[\tilde{\mathcal{S}},\alpha_i] -132 m_Q^5\langle\bar q q\rangle I_{ls}[2,1] I_4[\tilde{\mathcal{S}},\alpha_i]  \nonumber\\& 
	+66 m_Q^7\langle\bar q q\rangle I_{ls}[3,1] I_4[\tilde{\mathcal{S}},\alpha_i] +33f_{3\gamma}m_Q^4 I_{ls}[2,2] I_4[\mathcal{V},\alpha_i] -33f_{3\gamma}m_Q^8 I_{ls}[4,2] I_4[\mathcal{V},\alpha_i] \nonumber\\&
	-12m_Q^5\langle\bar q q\rangle I_{ls}[2,1] I_5[\mathbb{A},u] + 12 m_Q^7\langle\bar q q\rangle I_{ls}[3,1]
	I_5[\mathbb{A},u] +12\chi m_Q^5\langle\bar q q\rangle I_{ls}[2,2] I_5[\phi_\gamma,u] \nonumber\\& 
	-24\chi m_Q^7\langle\bar q q\rangle  I_{ls}[3,2] I_5[\phi_\gamma,u] +12 \chi m_Q^9\langle\bar q q\rangle  I_{ls}[4,2] I_5[\phi_\gamma,u] +12 f_{3\gamma}m_Q^6 I_{ls}[3,2] I_5[\psi^a,u] \nonumber\\& 
	-12 f_{3\gamma}m_Q^8 I_{ls}[4,2] I_5[\psi^a,u] \Bigg\} \nonumber\\&
	-\frac{1}{2304\,M^{2}} e^{-\frac{m_Q^2}{M^{2}}} e_q\, \langle g_s^2G^2 \rangle
	\Bigg\{m_Q\left(m_Q^2+M^{2}\right) \langle\bar q q\rangle \mathbb{A}[u_0] -4f_{3\gamma}m_Q^2M^{2}\psi^a[u_0]
	\nonumber\\&
	+2m_QM^{2}\langle\bar q q\rangle I_1[h_\gamma,u] +4m_QM^{2}\langle\bar q q\rangle
	I_2[h_\gamma,u] +22\,m_QM^{2}\langle\bar q q\rangle I_3[\mathcal{S},\alpha_i] \nonumber\\&
	-11\,m_QM^{2}\langle\bar q q\rangle I_3[\mathcal{T}_1,\alpha_i]
	-11\,m_QM^{2}\langle\bar q q\rangle I_3[\mathcal{T}_2,\alpha_i]
	+11\,m_QM^{2}\langle\bar q q\rangle I_3[\mathcal{T}_3,\alpha_i] \nonumber\\&
	+11\,m_QM^{2}\langle\bar q q\rangle I_3[\mathcal{T}_4,\alpha_i] +22\,m_QM^{2}\langle\bar q q\rangle
	I_3[\tilde{\mathcal{S}},\alpha_i]\Bigg\} \nonumber\\& - \frac{1}{2304\,M^{2}} e^{-\frac{m_Q^2}{M^{2}}}
	e_q\,\langle g_s^2G^2 \rangle\,I_s[0] \Bigg\{4\, \chi\, e^{\frac{m_Q^2}{M^{2}}} m_QM^{2}\langle\bar q q\rangle \varphi_\gamma[u_0] \nonumber\\& -12e^{\frac{m_Q^2}{M^{2}}} f_{3\gamma}M^{2}\psi^a[u_0] +4e^{\frac{m_Q^2}{M_{\rm sq}}} f_{3\gamma}M^{2}\psi^{v}[u_0] -8e^{\frac{m_Q^2}{M^{2}}} f_{3\gamma}M^{2} I_1[\psi^{v},u] \nonumber\\&
	+11e^{\frac{m_Q^2}{M^{2}}} f_{3\gamma}M^{2} I_4[\mathcal{A},\alpha_i] +11e^{\frac{m_Q^2}{M^{2}}}
	f_{3\gamma}M^{2} I_4[\mathcal{V},\alpha_i] +2e^{\frac{m_Q^2}{M^{2}}} f_{3\gamma}M^{2} I_5[\psi^a,u] \Bigg\}\,,
\end{align}
while the second, $\Pi^{\prime \rm QCD}(\rm{M^2},\rm{s_0})$, reads:
\begin{align}\label{CorrFnonpertQCD}  
	\Pi^{\prime \mathrm{QCD}}&(\rm{M^2},\rm{s_0}) = \frac{e_q g_s^2 m_q^2 m_Q^4}{64\pi^4}
	\left(I_{ls}[3,1]-m_Q^2I_{ls}[4,1] \right) + \frac{1}{576\,M^{2}} e^{-\frac{m_Q^2}{M^{2}}} e_q\,\langle g_s^2G^2 \rangle \Bigg[ 4f_{3\gamma}M^{2} I_1[\psi^{v},u] \nonumber\\&
	-11\Bigg(f_{3\gamma}M^{2} I_3[\mathcal{A},\alpha_i] +\,m_Q\langle\bar q q\rangle
	\Big( I_3[\mathcal{T}_1,\alpha_i]+I_3[\mathcal{T}_2,\alpha_i] -I_3[\mathcal{T}_3,\alpha_i] -I_3[\mathcal{T}_4,\alpha_i]\Big)\Bigg)\Bigg] \nonumber\\& + \frac{e_q\,m_q^2}{384\pi^2}\, \langle g_s^2G^2 \rangle\, e^{-m_Q^2/M^2} \nonumber\\& 
	+ \frac{e_q g_s^2 m_Q}{384\pi^2} \Bigg\{-16 f_{3\gamma}\,m_Q^3 \Big(-I_{ls}[3,1]+ m_Q^2 I_{ls}[4,1]\Big) I_1[\psi^v,u] -44f_{3\gamma}\,m_Q^3 \Big(-I_{ls}[3,1]+m_Q^2I_{ls}[4,1]\Big) \nonumber\\&
	I_3[\mathcal A,\alpha_i]
	+11 \Bigg( \langle \bar qq \rangle \Big(3I_{ls}[1,0]-2m_Q^2I_{ls}[2,0]-m_Q^4I_{ls}[3,0]\Big)
	I_3[\mathcal T_1,\alpha_i] \nonumber\\&
	+\langle \bar qq \rangle \Big(3I_{ls}[1,0]-2m_Q^2I_{ls}[2,0]-m_Q^4I_{ls}[3,0]\Big)
	I_3[\mathcal T_2,\alpha_i] 
	-3 \langle \bar qq \rangle \,I_{ls}[1,0] I_3[\mathcal T_3,\alpha_i] \nonumber\\&
	+2m_Q^2 \langle \bar qq \rangle \,I_{ls}[2,0] I_3[\mathcal T_3,\alpha_i] 
	+m_Q^4 \langle \bar qq \rangle \,I_{ls}[3,0] I_3[\mathcal T_3,\alpha_i]
	-3 \langle \bar qq \rangle \,I_{ls}[1,0] I_3[\mathcal T_4,\alpha_i] \nonumber\\&
	+2m_Q^2 \langle \bar qq \rangle \,I_{ls}[2,0]I_3[\mathcal T_4,\alpha_i]
	+m_Q^4 \langle \bar qq \rangle \,I_{ls}[3,0]I_3[\mathcal T_4,\alpha_i] 
	- f_{3\gamma} \,m_Q I_{ls}[2,1] I_4[\mathcal A,\alpha_i] \nonumber\\&
	+2 f_{3\gamma} \,m_Q^3 I_{ls}[3,1] I_4[\mathcal A,\alpha_i] 
	-f_{3\gamma} \,m_Q^5 I_{ls}[4,1] I_4[\mathcal A,\alpha_i]\Bigg) \Bigg\}\,,
\end{align}
where the following integrals are used in the above expressions:
\begin{align}\label{DAsinteg}
	I_{ls}[n,m]&= \int_{(m_q+m_Q)^2}^{s_0} ds \int_{(m_q+m_Q)^2}^s dl~ e^{-s/M^2}~\frac{(s-l)^m}{l^n}\nonumber\,, \\
	I_s[n]&=  \int_{(m_q+m_Q)^2}^{s_0} ds~e^{-s/M^2}~s^n~,\nonumber\\
	I_1[\mathcal{DA},u]&=\int_0^1 du~ \mathcal{DA}(u),\nonumber\\
	I_2[\mathcal{DA},u]&=\int_0^1 du~ u\,\mathcal{DA}(u),\nonumber\\
	I_3[\mathcal{DA},\alpha_i]&=\int D_{\alpha_i} \int_0^1 dv~ \mathcal{DA}(\alpha_{\bar q},\alpha_q,\alpha_g)\,
	\delta(\alpha_ q +(1-v) \alpha_g-u_0),\nonumber\\	
	I_4[\mathcal{DA},\alpha_i]&=\int D_{\alpha_i} \int_0^1 dv~ \mathcal{DA}(\alpha_{\bar q},\alpha_q,\alpha_g)\,
	\delta^\prime (\alpha_ q +(1-v) \alpha_g-u_0),\nonumber\\	
	I_5[\mathcal{DA}]&=\int_0^1 du~ u\, \mathcal{DA}(u)\, \delta'(u-u_0)\,.
\end{align}

\bibliographystyle{JHEP}
\bibliography{bibliography}

\end{document}